\documentclass[12pt]{article}
\usepackage{bbm}
\usepackage{graphics, color, amsfonts, dsfont, graphicx, amssymb, epsfig}
\usepackage{amsmath,amssymb}
\usepackage{mathtools}
\usepackage{amsthm}
\usepackage{mathrsfs}
\usepackage{colortbl}
\usepackage{times}
\usepackage{geometry}
\usepackage{listings}
\usepackage{xcolor}
\usepackage{color}
\usepackage{bm}
\usepackage{hyperref}
\usepackage{multirow}
\usepackage{algpseudocode}
\usepackage{verbatim}
\usepackage{chngcntr}
\usepackage{booktabs}
\usepackage{tabularx}
\usepackage{fullpage, setspace}
\usepackage{makecell}
\usepackage{xr}
\usepackage{chngcntr} 
\usepackage{booktabs}
\usepackage{tabularx} 
\usepackage{fullpage, setspace}
\usepackage{makecell}
\usepackage{xr}
\usepackage{siunitx}
\usepackage{caption}
\usepackage{booktabs,subcaption,amsfonts,dcolumn}

\newtheorem{theorem}{Theorem}
\newtheorem{assumption}{Assumption}

\newtheorem{proposition}{Proposition}  
\newtheorem{lemma}{Lemma} 

\newtheorem{remark}{Remark}
\usepackage[linesnumbered,ruled,vlined]{algorithm2e}
\SetKwInput{KwInput}{Input}                
\SetKwInput{KwOutput}{Output}              
\SetKw{Break}{break}                      
\usepackage{natbib}
\usepackage[nottoc]{tocbibind}
\newcommand{\diag}{\text{diag}}
\newcommand{\rank}{\text{rank}}
\newcommand{\tr}{\text{tr}}

\pdfoutput=1
\begin{document}
	
	\title{Change point detection in dynamic heterogeneous networks via subspace tracking}
\author{Yuzhao Zhang$^{\dag}$$^\ddag$, Jingnan Zhang$^{\S}$, Yifan Sun$^{\dag}$ and Junhui Wang$^\P$\\ [10pt]
	$^\dag$Center for Applied Statistics and School of Statistics\\
	Remin University of China
	\and 
	$^\ddag$School of Data Science \\
	City University of Hong Kong 
	\and
	$^\S$International Institute of Finance, School of Management\\
	University of Science and Technology of China 
	\and
	$^\P$Department of Statistics \\
	The Chinese University of Hong Kong }
	\date{ }
	
	\maketitle
	
	\onehalfspacing
	\begin{abstract}
		Dynamic networks consist of a sequence of time-varying networks, and it is of great importance to detect the network change points. Most existing methods focus on detecting abrupt change points, necessitating the assumption that the underlying network probability matrix remains constant between adjacent change points. This paper introduces a new model that allows the network probability matrix to undergo continuous shifting, while the latent network structure, represented via the embedding subspace, only changes at certain time points. Two novel statistics are proposed to jointly detect these network subspace change points, followed by a carefully refined detection procedure. Theoretically, we show that the proposed method is asymptotically consistent in terms of change point detection, and also establish the impossibility region for detecting these network subspace change points. The advantage of the proposed method is also supported by extensive numerical experiments on both synthetic networks and a UK politician social network.
	\end{abstract}
	
	
	
	\doublespacing
	
	\section{Introduction}
	\label{sec:intro}
	
	Network provides a versatile framework for depicting pairwise interactions among various entities, such as social networks \citep{barabasi2002evolution, Chen2021}, biological networks \citep{Voytek2015, Ozdemir2017}, and economical networks \citep{page2005networks, farrell2019weaponized}. When interactions among entities are documented with time stamps, it gives rise to dynamic networks, where one of the central challenges is to detect time points when network structure changes substantially. 
	
\begin{figure}[!htb]
		\centering
		\begin{subfigure}[b]{0.5\textwidth}
			\centering
			\includegraphics[width=\textwidth]{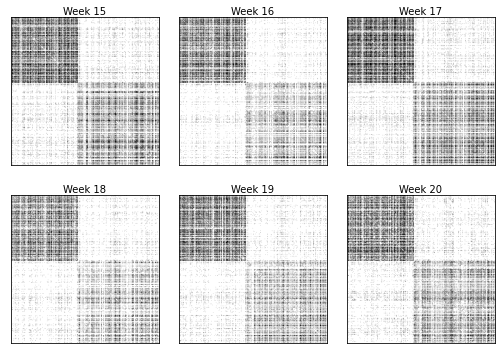} 
			\label{fig:prob1}
		\end{subfigure}
		\hfill
		\begin{subfigure}[b]{0.45\textwidth}
			\includegraphics[width=\textwidth, height=2.2in]{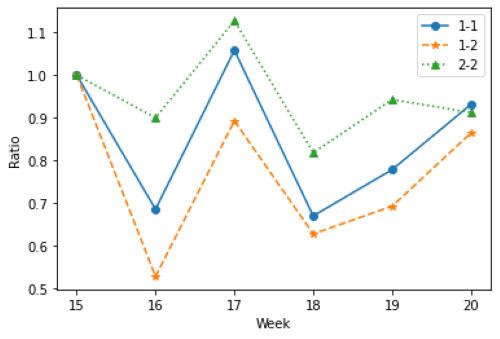}
			\label{fig:prob2}
		\end{subfigure}
		\caption{The left panel displays the adjacency matrices of 516 UK politicians over 6 weeks, where the structure with two communities is clear and stable. We then fit a multi-layer SBM model \citep{Lei2022} and obtain the underlying network probability matrix $\bm{\widehat P}_t$ for each week. The right panel displays the ratios between each element of $\bm{\widehat P}_t$ over that of the first week. It is clear that $\bm{\widehat P}_t$ changes substantially from week to week. }    
		\label{fig:prob}
	\end{figure}

	Extensive research has been dedicated to detecting change points in dynamic networks. A commonly adopted approach is to simply convert networks into high-dimensional vectors and then apply classical vector-based change points detection algorithms, such as \citet{Bhattacharyya2020}, \citet{Wang2021} and \citet{Chen2021}. Another popular route assumes some specific network generative models, such as the random dot product model \citep{Padilla2022}, the graphon-based model \citep{Zhao2019} or the stochastic block model (SBM; \cite{Monika2020, Xu2022}). In all the aforementioned works, the underlying network probability matrices are assumed to remain unchanged between any two adjacent change points, and thus piece-wise constant over time. This assumption can be stringent in many real-world applications, since the network probability matrices may change continuously, even though the network structure often remains unchanged for a period of time \citep{Weaver2018,Ozdemir2017,Cheung2020,Zhang2020online}. For instance, Figure \ref{fig:prob} displays the social networks among 516 UK politicians over 6 weeks \citep{Weaver2018}, where the community formations are stable but the network connection probabilities appear to be fairly versatile. 
	
	In this paper, we consider dynamic networks with continuously changing connection probabilities, where the change points are defined as those when network structure changes substantially. In literature, only a few approaches \citep{Cribben2017,Ozdemir2017,Cheung2020} are developed to detect such change points in dynamic networks, yet most are algorithm-based and lack theoretical justification. Specifically, \citet{Cheung2020} and \citet{Cribben2017} need to estimate the community membership over a large number of sub-intervals, leading to heavy computational burden; and \citet{Ozdemir2017} requires several stringent and rather unrealistic constraints on the evolution of network subspace, which largely restricts its applicability. 
	
	We develop a change point detection method via subspace tracking, where each network layer is embedded in a low-dimensional subspace, and the network change points are detected by tracking the changes of the embedding subspace. Particularly, we introduce two novel detection statistics to jointly detect the subspace change points, one focusing on the bases of the subspaces and the other on their ranks. A refined detection algorithm is built upon these two detection statistics as well as some fine tuning to attain both theoretical guarantees and superior numerical performance in detecting the network subspace change points. To the best of our limited knowledge, the proposed method is the first network subspace change point detection method with theoretical guarantees. More convincingly, we also establish the impossibility region for detecting the network subspace change points for a minimax perspective, which essentially demonstrates that the proposed method is theoretically optimal up to a logarithm factor.

	The rest of the paper is organized as follows. Section 2 presents the proposed method for detecting network subspace change points, and elucidates the detection algorithms. Section 3 establishes the asymptotic consistency of our proposed method and establishes the impossibility region for detecting the network subspace change points. Section 4 examines the numerical performance of the proposed method on both synthetic and real-world networks, and compares it against some existing competitors. A brief discussion is contained in Section 5, and the technical proofs are contained in the Appendix. The auxiliary lemmas can be found in the Supplementary Materials.
	
	\paragraph{Notations.} For any positive integer $K$, denote $\{1,\ldots, K\}$ as $[K]$ for short. For any sequences $\{a_n\}_{n=1}^{\infty}$ and $\{b_n\}_{n=1}^{\infty}$, we write $a_n \prec b_n$ if $a_n=o(b_n)$ and $a_n \preceq b_n$ if $a_n=O(b_n)$. Further, if $a_n = O(b_n)$ and $b_n=O(a_n)$, we write $a_n \simeq b_n$. Denote by $e_{k}(K)$ the $k$-th standard orthonormal basis of $\mathbb{R}^{K}$, and $\bm{1}_{K}$ the $K$-dimensional vector of ones. For a square matrix $\bm{A} \in \mathbb{R}^{n \times n}$, its rank is denoted as $\rank(\bm{A})$, its trace is denoted as $\tr(\bm A)$, and its eigenvalues can be sorted in a descending order $\lambda_{\max}\big(\bm{A}\big) = \lambda_{1}\big(\bm{A}\big)\geq\lambda_{2}\big(\bm{A}\big) \geq \cdots \geq \lambda_{n}\big(\bm{A}\big) = \lambda_{\min}\big(\bm{A}\big)$. Similarly, we also sort its singular value in a descending order $\sigma_{\max}\big(\bm A\big) = \sigma_{1}\big(\bm A\big) \geq \sigma_{2}\big(\bm A\big) \geq \cdots \geq \sigma_{n}\big(\bm A\big) = \sigma_{\min}\big(\bm A\big)$. Let $\bm{A}_{S,S'}$ denote a submatrix of $\bm{A}$ with row indices in $S$ and column indices in $S'$. For an orthonormal matrix $\bm{V} \in \mathbb{R}^{n \times m}$ with $n>m$, a basis of its orthogonal complement is denoted as $\bm{V}_{\perp}$. 
	
	\section{Proposed method}
	\label{section:subspace-model}
	
	For dynamic network $\{{\cal G}_{t}\}_{t=1}^T$, its symmetric adjacency matrix for each network layer is denoted as $\bm{A}_{t} \in \{0,1\}^{n \times n}$ with $\bm{P}_{t} = E(\bm{A}_{t}) \in [0,1]^{n\times n}$. We consider a low-rank network embedding model \citep{Arroyo2021}, 
	$$
	\bm{P}_t = \rho_{n,T}\bm{V}_{t}\bm{M}_{t}\bm{V}_{t}^{\top},
	$$
	where $\rho_{n,T} = \max_{t \in [T]}\|\bm P_{t}\|_{\max}$ controls the network sparsity that may vanish with $n$ and $T$, $\bm{M}_{t} \in \mathbb{R}^{R \times R}$ is a full-rank matrix, and $\bm{V}_t \in \mathbb{R}^{n \times R}$ is an orthonormal matrix. Apparently, $\bm{P}_{t}$ lies in the subspace spanned by the columns of $\bm{V}_{t}$, and it is identifiable up to an orthogonal transformation due to the full-rank $\bm{M}_t$.
	
	Suppose there are a total of $K$ change points in $\{\bm{P}_{t}\}_{t=1}^{T}$, denoted as $\Gamma = \{\tau_{k}\}_{k=1}^{K}$ with $1 = \tau_0 < \tau_1 < \cdots < \tau_{K} < \tau_{K+1}=T$, where 
	\begin{align}
		{\bm P}_{t}=\rho_{n,T}\bm{V}^{(k)}\bm{M}_t(\bm{V}^{(k)})^{\top},
		\label{eq:subspace-model}
	\end{align} 
	for any $t \in[\tau_{k-1},\tau_{k})$, and $\min_{k\in[K]}\|\bm{V}^{(k)}(\bm{V}^{(k)})^{\top}-\bm{V}^{(k+1)}(\bm{V}^{(k+1)})^{\top}\|_{F}>0$. It is important to remark that ${\bm P}_{t}$ may change continuously over time, whereas the latent subspace defined by $\bm V$ can only change at each $\tau_k$. A special case of \eqref{eq:subspace-model} has been considered in \citet{Cheung2020}, assuming $\bm{P}_{t}=\rho_{n,T}\bm{Z}^{(k)}\bm{B}_{t}(\bm{Z}^{(k)})^{\top}$ for any $t\in[\tau_{k-1},\tau_{k})$, where $\bm{Z}^{(k)} \in \{0,1\}^{n \times R}$ denotes the community membership. We term the change points defined in \eqref{eq:subspace-model} as the network subspace change points, which are in sharp contrast to the network probability change points in literature \citep{Wang2017,Monika2020,Wang2021}, which assumes that ${\bm P}_{t}$ can only change at each $\tau_k$ and otherwise remains constant within each $[\tau_{k-1}, \tau_k)$.

	\subsection{Two detection statistics}
	
	We first introduce two statistics to jointly detect the subspace change points of $\{\bm{P}_{t}\}_{t=1}^{T}$. The following auxiliary lemma is necessary.
	
	\begin{lemma}
		\label{lemma:1}
		Let $\bm U \in \mathbb{R}^{n\times R_{u}}$ and $\bm V\in \mathbb{R}^{n\times R_{v}}$ be two orthognormal matrices with $R_{u}+R_{v}\prec n$. It holds true for any $\bm{Q}\in \mathbb{R}^{R_{v} \times R_{v}}$ that
		\begin{align*}
			\|{\bm U}_{\perp}^{\top} \bm{V}\|_{2}^2 & \geq \big(R_v - R_u + \|\bm{U} \bm{U}^{\top}-\bm{V} \bm{V}^{\top} \|_{F}^2\big)/(2R_{v}), \\
			\big\|{\bm U}_{\perp}^{\top} {\bm V} \bm{Q} \bm{V}^{\top}\big\|_{2} & \geq \sigma_{\min}\big(\bm{Q}\big)\sqrt{\big(R_v - R_u + \|\bm{U} \bm{U}^{\top}-\bm{V} \bm{V}^{\top}\|_{2}^2\big)/(2R_{v})}.
		\end{align*}
	\end{lemma}
	
	Let $\Delta_{\min}= \min_{k \in [K+1]} (\tau_k-\tau_{k-1})$ denote the minimum distance between two adjacent change points. We examine the behavior of $\sum_{t=l-L+1}^{l} \bm{P}_{t}^2$ and  $\sum_{t=l}^{l+L-1} \bm{P}_{t}^2$  with  $L\leq\Delta_{\min}/3$, where summing up squared probability matrices has been proven to be effective in summarizing signals from multiple layers \citep{Lei2022}. Particularly, for any $ \tau_{k-1}+2L \le l < \tau_{k}+2L$, the first detection statistic measures the signal strength of projecting $\sum_{t=l-L+1}^{l}\bm{P}_{t}^2$ onto $\bm{V}_{\perp}^{(k)}$, defined as 
	\begin{align}
		\Pi_{proj}(l) = \Big\|(\bm{V}_{\perp}^{(k)})^{\top}\sum_{t=l-L+1}^{l}\bm{P}_{t}^2\Big\|_{2}.
		\label{eq:def-proj}
	\end{align}
	It behaves differently before and after $\tau_k$, as long as  $\rank (\bm{V}^{(k+1)}) \ge \rank(\bm{V}^{(k)})$. Note that the scenario with $\rank (\bm{V}^{(k+1)}) < \rank( \bm{V}^{(k)})$ also indicates a clear network subspace change point. Thus, with $R^k= \rank(\bm{V}^{(k)})$, the second statistic is defined as 
	\begin{align}
		\Pi_{eig}(l) = \lambda_{R^{(k)}}\Big(\sum_{t=l}^{l+L-1}\bm{P}_{t}^2\Big).
		\label{eq:def-eig}
	\end{align} 
	
	These two detection statistics can work jointly to detect the network subspace change points in $\{\bm{P}_{t}\}_{t=1}^T$. Specifically, when $\tau_{k-1}+2L \le l <\tau_{k}$, we have $\sum_{t=l-L+1}^l \bm{P}_{t}^2 =\sum_{t=l-L+1}^l \bm{V}^{(k)} \bm{M}_{t}^2 (\bm V^{(k)})^{\top}$, which lies in the column space of $\bm V^{(k)}$ and thus is orthogonal to $\bm V^{(k)}_\perp$. Moreover, as $\bm{M}_{t}$ is full rank, we have
	$$
	\lambda_{R^{(k)}}\Big(\sum_{t=l}^{l+L-1} \bm{P}_{t}^2\Big) \geq \lambda_{R^{(k)}}\Big(\sum_{t=l}^{\min\{l+L-1,\tau_{k}-1\}} \bm{M}_{t}^2\Big)>0,
	$$ 
	It thus holds true that $\Pi_{proj}(l) =0$ and $\Pi_{eig}(l)>0$ when $\tau_{k-1}+2L \le l<\tau_{k}$. When $\tau_{k}\leq l < \tau_{k}+2L$, it follows from Lemma \ref{lemma:1} that
	\begin{align*}
		\Pi_{proj}(l) & = \Big\|(\bm{V}_{\perp}^{(k)})^{\top}\sum_{t=l-L+1}^{l} \bm{P}_{t}^2\Big\|_{2} = \rho_{n,T}^2\Big\|(\bm{V}_{\perp}^{(k)})^{\top}\bm{V}^{(k+1)}\sum_{t=\max\{\tau_{k},l-L+1\}}^{l} \bm{M}_{t}^2(\bm{V}^{(k+1)})^{\top}\Big\|_{2}\notag \\
		& \geq \rho_{n,T}^2\lambda_{\min}\Big(\sum_{t=\max\{\tau_{k},l-L+1\}}^{l} \bm{M}_{t}^2 \Big)\sqrt{(R^{(k+1)}-R^{(k)}+\delta^{k+1})/(2R^{(k+1)})}.
	\end{align*} 
	It immediately implies that $\Pi_{proj}(l)>0$ as long as $R^{(k+1)}-R^{(k)} \ge 0$. 
	On the other hand, if $R^{(k)}>R^{(k+1)}$, then 
	$$
	\Pi_{eig}(l) = \lambda_{R^{(k)}}\Big(\sum_{t=l}^{l+L-1} \bm{P}_{t}^2\Big)=\lambda_{R^{(k)}}\Big(\bm{V}^{(k+1)}\sum_{t=l}^{l+L-1} \bm{M}_{t}^2(\bm{V}^{(k+1)})^{\top}\Big)=0. 
	$$
	Therefore, a time point $l$ is a network subspace change point if $\Pi_{proj}(\cdot)$ rises from $0$ at $l$ or $\Pi_{eig}(\cdot)$ drops to $0$ at $l$.
	
\begin{figure}[!htb]
		\centering
		\includegraphics[width=0.6\hsize]{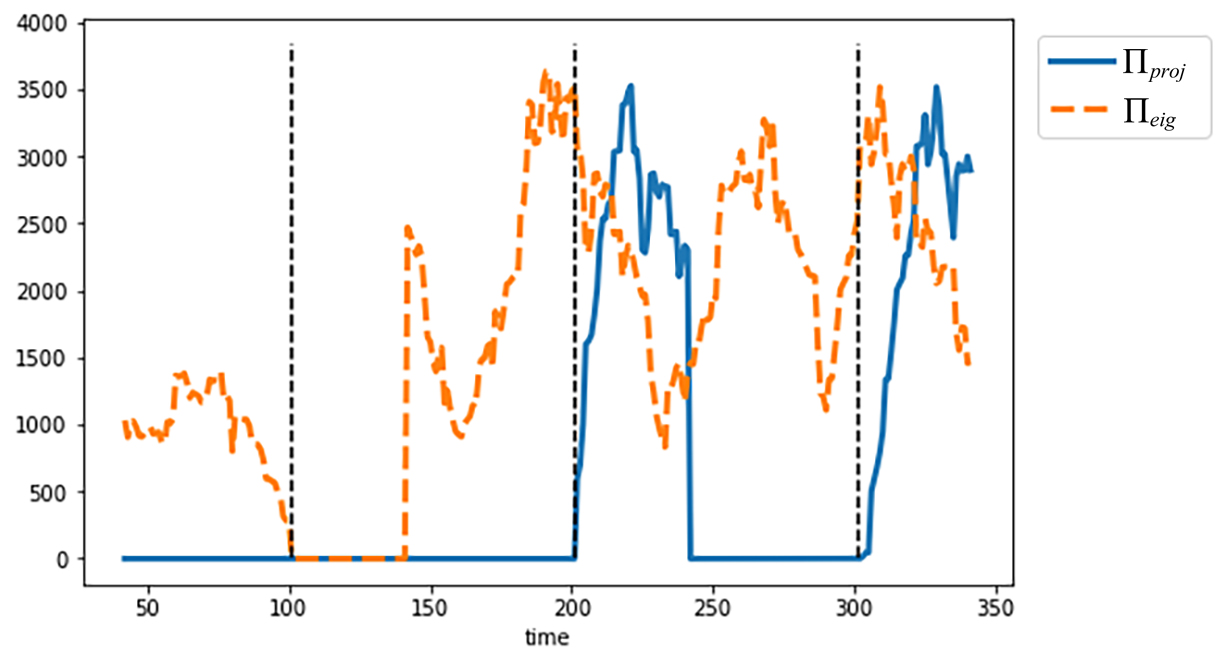}
		\caption{The curves of $\Pi_{proj}(l)$ and $\Pi_{eig}(l)$ for $l \in [41,340]$, and the dashed vertical lines mark the true change points at 101, 201, and 301.  }
		\label{fig:toy}
	\end{figure}

	We now illustrate the utility of the two statistics for jointly detecting the network subspace change points in a toy example with $n=100$, $T=400$, and $\Gamma = \{101,201,301\}$. We generate $\bm{V}^{(1)}, \ldots, \bm{V}^{4}$ such that $\rank(\bm{V}^{(1)}) > \rank(\bm{V}^{(2)}) <\rank(\bm{V}^{(3)}) = \rank(\bm{V}^{(4)})$, so that the subspace ranks decrease, increase or remain unchanged at the three change points, respectively. We simply set $L=20$ and calculate the two statistics over the time interval $[41,340]$. The values of both statistics are summarized in Figure \ref{fig:toy}. It is clear that $\Pi_{eig}(l)$ drops towards 0 at $l=101$, and $\Pi_{proj}(l)$ rises away from 0 after $l=201$ and $301$, respectively. Therefore all three change points can be perfectly identified by examining the behavior of $\Pi_{proj}(l)$ and $\Pi_{eig}(l)$. 
	
	\subsection{Detection algorithm}
	\label{sec:algorithm}
	
	The two proposed detection statistics in Section 2.1 lead to a simple recursive algorithm for detecting network subspace change points. Particularly, we first set $\widetilde{\tau}_{0} = 1$. Given $\widetilde{\tau}_{k-1}$, for $l = \widetilde{\tau}_{k-1}+2L, \widetilde{\tau}_{k-1}+2L+1, \ldots$, we compute  
	\begin{align}
		\widehat{\Pi}_{proj}(l) = \Big\|(\widehat{\bm V}^{(k)}_{\perp})^{\top}\sum_{t=l-L+1}^{l}\big(\bm{A}_{t}^2-\bm{D}_{t}\big)\Big\|_{2} \ \  \mbox{and} \ \
		\widehat{\Pi}_{eig}(l) = \lambda_{\rank(\widehat{\bm V}^{(k)})} \bigg( \sum_{t=l}^{l+L-1}\bm{A}_{t}^2-\bm{D}_{t}\bigg ).
		\label{eq:detect-statistics}
	\end{align}
	In both detection statistics, the unknown term $\bm{P}_{t}^2$ is replaced by its estimate $\bm{A}_{t}^2 - \bm{D}_t$ with $\bm{D}_{t} = \diag(\bm{A}_{t}\mathbf{1}_{n})$ \citep{Lei2022}, and $\widehat{\bm V}^{(k)}$ contains the leading eigenvectors of $\sum_{t=\widetilde{\tau}_{k-1}+L}^{\widetilde{\tau}_{k-1}+2L-1}(\bm{A}_{t}^2-\bm{D}_{t})$ with eigenvalues larger than a pre-specified thresholding value $b_{n,T}$. The first $l$ such that $\widehat{\Pi}_{proj}(l) > (1+\sqrt{2})b_{n,T}$ or $\widehat{\Pi}_{eig}(l) < b_{n,T}$ is set as the next estimated change point $\widetilde{\tau}_{k}$, and we repeat the above procedure for $l = \widetilde{\tau}_k+2L, \widetilde{\tau}_k+2L+1, \ldots$. The whole procedure continues until $l$ reaches $T-L$, and the resultant estimated change points are denoted as $\widetilde{\Gamma} = \{\widetilde{\tau}_{k}\}_{k=1}^{\widetilde{K}}$.

	Note that these estimates rely heavily on the pre-specified threshold $b_{n, T}$, and thus an appropriate $b_{n,T}$ may yield sub-optimal detection of the network subspace change points in practice if the chosen $b_{n,T}$ is not suitable. We propose to further refine the detection algorithm, to be more robust against the choice of $b_{n,T}$. A universal eigenvalue threshold operator is defined as
	$$\text{UEVT}\big(\bm{G},h\big) = \sum_{i:\lambda_{i}(\bm{G})>h}\lambda_{i}\big(\bm{G}\big) \bm{u}_{i} \bm{u}_{i}^\top,$$
	where $\bm{G}=\sum_{i=1}^{n}\lambda_{i}\big(\bm{G}\big) \bm{u}_{i} \bm{u}_{i}^\top$ denotes the standard eigen-decomposition of $\bm{G}$, and $h$ is a generic thresholding value. The UEVT operator constructs a low-rank approximation of $\bm{G}$ by excluding small eigenvalues and their associated eigenvectors.  We consider a low-rank estimate for $\sum_{t=l-L+1}^{l}\bm{P}_{t}^2$,
	$$
	\bm{\bar P}^{(l)} = \text{UEVT}\Big( \sum_{t=l-L+1}^{l} (\bm{A}_{t}^2 - \bm{D}_t), b_{n,T} \Big),	
	$$ 	
	whose rank is the cardinality of $\big\{i: \lambda_{i}\big(\sum_{t=l-L+1}^{l} (\bm{A}_{t}^2-\bm{D}_{t}) \big)>b_{n,T}\big\}$. Then we introduce two additional statistics,
	\begin{align}
		\widehat{\Pi}_{ref1}(l)=\tr\big(\widehat{\bm U}_{\perp}^{(l-1)}(\widehat{\bm U}_{\perp}^{(l-1)})^{\top}\bar{\bm P}^{(l+L-1)}\big) \text{, and } \widehat{\Pi}_{ref2}(l)=\tr\big(\widehat{\bm U}_{\perp}^{(l+L-1)}(\widehat{\bm U}_{\perp}^{(l+L-1)})^{\top}\bar{\bm P}^{(l-1)}\big),
		\label{eq:refine-statistics}
	\end{align}
	where $\widehat{\bm U}^{(l-1)}$ and $\widehat{\bm U}^{(l+L-1)}$ contain the eigenvectors corresponding to the nonzero eigenvalues of $\text{UEVT}(\sum_{t=l-L}^{l-1}\bm{A}_{t}^2-\bm{D}_{t},b_{n,T})$ and  $\text{UEVT}(\sum_{t=l}^{l+L-1}\bm{A}_{t}^2-\bm{D}_{t},b_{n,T})$, respectively. 
	
	For each $k \in [\widetilde{K}]$, we can refine the estimation of the change points by maximizing $\widehat{\Pi}_{ref1}(l)$ or $\widehat{\Pi}_{ref2}(l)$ based on the comparison of $\widehat{R}^{(k)}$ and $\widehat{R}^{(k+1)}$ in some range of $l$. Specifically, if $\widehat{R}^{(k)}<\widehat{R}^{(k+1)}$, it can be shown that the real change point $\tau_{k}^{*}$ is contained in $[\widetilde{\tau}_{k}-L+1,\widetilde{\tau}_{k}]$ with high probability. Then we denote the refined estimate as $\widehat{\tau}_{k} = \arg\max_{l \in [\widetilde{\tau}_{k}-L+1,\widetilde{\tau}_{k}]}\widehat{\Pi}_{ref1}(l)$. If $\widehat{R}^{(k)}>\widehat{R}^{(k+1)}$, it can be shown that  $\tau_{k}^{*}$ is contained in $[\widetilde{\tau}_{k},\widetilde{\tau}_{k}+L-1]$ with high probability. Thus, we denote the refined estimate as $\widehat{\tau}_{k} = \arg\max_{l \in [\widetilde{\tau}_{k},\widetilde{\tau}_{k}+L-1]}\widehat{\Pi}_{ref2}(l)$. If $\widehat{R}^{(k)}=\widehat{R}^{(k+1)}$, we denote the first $l \in \{ \widetilde{\tau}_{k}-1,\widetilde{\tau}_{k}-2,\cdots\}$ satisfying $\|(\widehat{\bm V}_{\perp}^{(k+1)})^{\top}\sum_{t=l}^{l+L-1}(\bm{A}_{t}^2-\bm{D}_{t})\|_{2}>(1+\sqrt{2})b_{n,T}$ as $\check{\tau}_{k}$. It can be shown that $\tau_{k}^{*}$ is contained in $[\check \tau_{k},\widetilde \tau_{k}]$ with high probability. Therefore, we set the refined estimate as  $\widehat{\tau}_{k} = \arg\max_{l \in [\check{\tau}_{k},\widetilde{\tau}_{k}]}\widehat{\Pi}_{ref1}(l)$. The detailed detection algorithm is summarized in Algorithm \ref{alg}. 
	
	\begin{algorithm}[!ht]
		\KwInput{$b_{n,T}$ and $L$}
		Set $k = 0$, $\widehat \tau_{k}=0$ and $l = 2L+1$
		
		Set $\widehat{\bm V}_{\perp}^{(k+1)}$ as the eigenvectors of $\sum_{t=L+1}^{2L}(\bm{A}_{t}^2-\bm{D}_{t})$ with eigenvalues not larger than $b_{n,T}$
		
		\While{$l \leq T-L$ }
		{
			Calculate $\widehat \Pi_{proj}(l)$ and $\widehat \Pi_{eig}(l)$ by Eq.\eqref{eq:detect-statistics}
			
			\If{$\widehat \Pi_{proj}(l)> (1+\sqrt{2})b_{n,T}$ or $\widehat \Pi_{eig}(l) < b_{n,T}$}
			{
				Set $k=k+1$ and $\widetilde \tau_k = l$
				
				Obtain $\widehat{\bm V}_{\perp}^{(k+1)}$ by selecting the eigenvectors of $\sum_{t=\widetilde \tau_k+L}^{\widetilde \tau_k+2L-1}(\bm{A}_{t}^2-\bm{D}_{t})$ with eigenvalues not larger than $b_{n,T}$
				
				Set $l=\widetilde \tau_k+2L$
			}
			\Else
			{Set $l = l+1$}
		}
		\For{$m = 1,2,\cdots,\widehat{K}$}{
			\If{$\rank(\widehat{\bm{V}}^{m}) < \rank(\widehat{\bm{V}}^{m+1})$}
			{
				Denote $\widehat \tau_{m} = \arg\max_{s \in [\widetilde{\tau}_{m}-L+1,\widetilde{\tau}_{m}]}\widehat{\Pi}_{ref1}(s)$, where $\widehat{\Pi}_{ref1}(s)$ is defined in Eq.\eqref{eq:refine-statistics}
			}
			\If{$\rank(\widehat{\bm{V}}^{m}) > \rank(\widehat{\bm{V}}^{m+1})$}
			{
				Denote $\widehat \tau_{m} = \arg\max_{s \in [\widetilde{\tau}_{m},\widetilde{\tau}_{m}+L-1]}\widehat{\Pi}_{ref2}(s)$, where $\widehat{\Pi}_{ref2}(s)$ is defined in Eq.\eqref{eq:refine-statistics}
			}
			\If{$\rank(\widehat{\bm{V}}^{m}) = \rank(\widehat{\bm{V}}^{m+1})$}
			{
				\For{$\check{\tau}_{m}=\widetilde{\tau}_{m}-1,\widetilde{\tau}_{m}-2,\cdots,\widehat{\tau}_{m-1}+1$}{
					\If{$\big\|(\widehat{\bm V}_{\perp}^{m+1})^{\top}\sum_{t=\check{\tau}_{m}}^{\check{\tau}_{m}+L-1}(\bm{A}_{t}^2-\bm{D}_{t})\big\|_{2}>(1+\sqrt{2})b_{n,T}$}
					{\Break}
				}
				Denote $\widehat \tau_{m} = \arg\max_{s \in [\check{\tau}_{m},\widetilde{\tau}_{m}]}\widehat{\Pi}_{ref1}(s)$  
			}
		}
		\KwOutput{$\widetilde{\Gamma}=\{\widetilde{\tau}_1,\dots,\widetilde{\tau}_{\widetilde{K}}\}$ and $\widehat{\Gamma}=\{\widehat{\tau}_1,\dots,\widehat{\tau}_{\widehat{K}}\}$ with $\widehat{K} = \widetilde{K}$}
		\caption{Subspace change point detection in dynamic networks}
		\label{alg}
	\end{algorithm}

	\section{Theory}
	
	Let the true underlying probability matrices be $\{\bm{P}_{t}^{*}\}_{t=1}^{T}$, and the true change points be $\Gamma^{*} = \{\tau_{k}^{*}\}_{k=1}^{K^*}$ with $1=\tau_{0}^{*}<\tau_{1}^{*}<\cdots<\tau_{K^{*}}^{*}<\tau_{K^{*}+1}^{*}=T$. Then, for any $t\in[\tau_{k-1}^{*},\tau_{k}^{*})$, we have
	\begin{align*}
		{\bm P}_{t}^{*}=\rho_{n,T}\bm{V}^{*(k)}\bm{M}_t^{*}(\bm{V}^{*(k)})^{\top}.
	\end{align*}
	Denote the minimum distance $\Delta_{\min}^{*} = \min_{k\in[K^{*}+1]} (\tau_{k}^*-\tau_{k-1}^*)$, and the minimum change magnitude $\delta_{\min}^* =  \min_{k \in [K^*]} \delta^{*(k)}$ with $\delta^{*(k)} = \|\bm{V}^{*(k+1)}(\bm{V}^{*(k+1)})^{\top}-\bm{V}^{*(k)}(\bm{V}^{*(k)})^{\top}\|_{F}^{2}$. Note that $\delta_{\min}^*\le 2\max_{k}R^{*(k)}\preceq 1$, where $R^{*(k)} = \rank(\bm{V}^{*(k)})$. The following technical assumptions are made.
	
	\begin{assumption}
		\label{assumption:balanced-singularvalue}
		There exists a sequence $\alpha_{n}\preceq 1$ such that $\sigma_{\max}(\bm{M}_{t}^{*})\simeq \sigma_{\min}(\bm{M}_{t}^{*})\geq n\alpha_n$ for any $t \in [T]$.
	\end{assumption}
	
	Assumption \ref{assumption:balanced-singularvalue} assures that $\bm{M}_{t}^{*}$ possesses balanced singular values, and its minimum singular value is asymptotically lowered bounded by $n \alpha_n$, where $\alpha_n$ may diminish to zero with $n$. This assumption is weaker than some commonly employed assumptions in network literature \citep{Chen2018,Lei2022}, which essentially implies Assumption \ref{assumption:balanced-singularvalue} with $\alpha_n \simeq 1$. 
	
	\begin{assumption}
		\label{assumption:signal-to-noise}
		(1) For dynamic networks with sparsity $\rho_{n,T} \succ 1/n$, 
		$$
		n\rho_{n,T} \alpha_{n}^4 \Delta_{\min}^{*} \delta^*_{\min} \succ \big(\log(n+T)\big)^2.
		$$ 
		(2) For dynamic networks with sparsity $\rho_{n,T}\preceq 1/n$, 
		$$
		n^2\rho_{n,T}^2 \alpha_{n}^{4} \Delta_{\min}^{*} \delta^*_{\min} \succ \big(\log(n+T)\big)^2.
		$$
	\end{assumption}
	
	Assumption \ref{assumption:signal-to-noise} specifies the required signal-to-noise ratio for the proposed method. Particularly, when the dynamic network is sparse with $\rho_{n,T}\simeq \log(n+T)/n$ and $\alpha_{n}\simeq 1$, Assumption \ref{assumption:signal-to-noise} simplifies to $\Delta_{\min}^{*}\delta_{\min}^{*}\succ \log(n+T)$. Note that $\Delta_{\min}^{*}\delta_{\min}^{*}$ is a popular quantity to measure the difficulty of a change point detection problem \citep{Yu2020}. If we further require $\delta_{\min}^{*}\simeq 1$, then our proposed method only requires $\Delta_{\min}^*$ to be slightly larger than $\log(n+T)$, which is equivalent to the requirement in \citet{Wang2021} on network probability change point. When the dynamic network is extremely sparse with $\rho_{n,T}\simeq 1/n$ and $\alpha_{n}\simeq 1$, Assumption \ref{assumption:signal-to-noise} simplifies to $\Delta_{\min}^{*}\delta_{\min}^{*}\succ \big(\log(n+T)\big)^2$, which is slightly stronger compared with the requirement when $\rho_{n,T}\simeq \log(n+T)/n$.  In fact, our proposed method is still applicable even when $\rho_{n,T}\succ \log(n+T)/(n\sqrt{T})$, provided that $\alpha_{n}\simeq 1$, $\Delta_{\min}^{*}\simeq T$ and $\delta_{\min}^{*}\simeq 1$. This sparsity requirement matches up with some of the latest results in the literature of multi-layer stochastic block model \citep{Lei2020,Lei2022}. 
	
	\begin{proposition}
		\label{prop:noise-upper-bound}
		Suppose Assumption \ref{assumption:signal-to-noise} holds, then for any $l\in [T]\setminus[L]$ with $\Delta_{\min}^{*} \preceq L\le \Delta_{\min}^{*}/3$, there exists positive constants $c_1$, $c_2$, and $c_3$ such that with probability at least $1-c_1(n+T)^{-2}$,it holds true that
		\begin{align*}
			\Big\|\sum_{t=l-L+1}^{l}\big(\bm{A}_{t}^2-\bm{D}_{t}-(\bm{P}_{t}^{*})^2\big)\Big\|_2 & \leq \frac{b_{n,T}}{\sqrt{\log(n+T)}},
			\text{ and } \\
			\big\|\sum_{t=l-L+1}^{l}(\bm{P}_{t}^{*})^2-\bar{\bm P}^{(l)}\big\|_{F} & \leq 4\max_{k \in [K^{*}+1]}\sqrt{R^{*(k)}}b_{n,T}, 
		\end{align*}
		when $n+T$ is large enough, where $b_{n,T} = c_2\big(Ln\rho_{n,T}^2+L^{1/2}n\rho_{n,T}\sqrt{\max\{c_3,n\rho_{n,T}\}}\big)\log(n+T)$.
	\end{proposition}
	Proposition \ref{prop:noise-upper-bound} extends Theorem 3 and 5 in \citet{Lei2022} to the context of dynamic networks with change points. It demonstrates that both $\sum_{t=l-L+1}^{l}(\bm{A}_{t}^2-\bm{D}_{t})$ and $\bar{\bm P}^{(l)}$ are good estimates of $\sum_{t=l-L+1}^{l}(\bm{P}_{t}^{*})^2$ under the spectral norm, whose estimation errors are controlled by the threshold $b_{n,T}$. These error bounds pave the theoretical foundation for quantifying the asymptotic behavior of the proposed methods.
	
	\begin{theorem}
		\label{theorem:alg2}
		Suppose Assumption \ref{assumption:balanced-singularvalue} and all assumptions in Proposition \ref{prop:noise-upper-bound} hold, and $n\succeq T$. Then it holds true that
		\begin{align*}
			\max_{k\in[K^{*}]}\frac{|\widehat{\tau}_{k}-{\tau}_{k}^{*}|}{\Delta_{\min}^{*}} \stackrel{p}{\rightarrow} 0.
		\end{align*}
	\end{theorem}
	Theorem \ref{theorem:alg2} shows that the estimate $\widehat \Gamma$, with appropriate choices of $L$ and $b_{n,T}$, is a consistent estimate of $\Gamma^{*}$ in terms of both change point number and locations. Particularly, the larger $L$ is, the more network layers are integrated to calculate the detection statistics, yet it also increases bias in estimating $\widehat{\bm V}^{(k)}$, as the interval $[\widetilde \tau_{k}+L,\widetilde \tau_{k}+2L-1]$ may include networks from both sides of a change point $\tau_{k}^{*}$. Also, the value of $b_{n,T}$ depends on the network sparsity. When the networks are relatively dense with $n\rho_{n,T} \succ 1$, we shall set $b_{n,T}$ at the same magnitude of $L^{1/2}(n\rho_{n,T})^{3/2}\log(n+T)$; when the networks are extremely sparse with $n\rho_{n,T} \preceq 1$, $b_{n,T}$ shall be set at the same magnitude of $L^{1/2}n\rho_{n,T}\log(n+T)$. 
	
	Then, we establish an impossible regime in terms of $(\alpha_n, \rho_{n,T},\Delta_{\min}^{*},\delta_{\min}^*)$, where no change point detection algorithm is consistent in detecting $\{\tau_{k}^{*}\}_{k=1}^{K^*}$.
	\begin{theorem}
		\label{theorem:mini-max}
		Let $\mathcal{Q}$ denote the class consisting of all multivariate Bernoulli distribution of $\{\bm{A}_t\}_{t=1}^{T}$. For any $Q \in \mathcal{Q}$, its corresponding probability matrices $\{\bm{P}_{t}^{(q)}\}_{t=1}^{T}$ must conform to Assumption \ref{assumption:balanced-singularvalue}, and the associated parameters satisfy $n\rho_{n,T}^{(q)}(\alpha_{n}^{(q)})^{4}\Delta_{\min}^{(q)}\delta_{\min}^{(q)}\simeq(\log(n+T))^{-1}$. Then there exists a sufficiently large $n_{0}$ and a positive constant $c_4$ such that for all $n\geq n_{0}$
		\begin{align*}
			\inf_{\check{\Gamma}}\sup_{Q \in \mathcal{Q}}\mathbb{E}_{Q}\Big(\frac{H\big(\check{\Gamma},\Gamma(Q)\big)}{\Delta_{\min}^{(q)}}\Big) \geq c_4,
		\end{align*}
		where $\check{\Gamma}$ can be any change point estimate, $\Gamma(Q)$ denotes the true network subspace change points under $Q\in\mathcal{Q}$, and $H(\cdot,\cdot)$ denotes the Hausdorff distance,
		\begin{align*}
			H\big(\check{\Gamma},\Gamma(Q)\big) = \max \Big \{\max_{\tau^{1}\in \Gamma(Q)}\min_{\tau^{2}\in \check{\Gamma}}|\tau^{1}-\tau^{2}|, \ \max_{\tau^{2}\in \check{\Gamma}}\min_{\tau^{1}\in \Gamma(Q)}|\tau^{1}-\tau^{2}| \Big \}.
		\end{align*}
	\end{theorem}
	
	Theorem \ref{theorem:mini-max} shows that, from a minimax perspective, there is no consistent change point detection algorithm if the parameters of the true underlying probability matrices $\{\bm{P}_{t}^{*}\}_{t=1}^{T}$ satisfy $n\rho_{n,T}\alpha_{n}^{4}\Delta_{\min}^{*}\delta_{\min}^{*}\preceq (\log(n+T))^{-1}$. It is interesting to remark that the proposed method is asymptotically consistent when $n\rho_{n,T}\succeq 1$ and  $n\rho_{n,T}\alpha_{n}^{4}\Delta_{\min}^{*}\delta_{\min}^{*}\succ \big(\log(n+T)\big)^2$, which is optimal up to a logarithm factor in view of the impossible regime above. 
	
	\begin{remark}
		When $n\rho_{n,T}\prec 1$, Assumption \ref{assumption:signal-to-noise} shall require an additional $(n\rho_{n,T})^{-1}$ term than the impossibility regime, which is possibly due to inherent challenge of estimating  $\sum_{t=l-L+1}^{l}(\bm{P}_{t}^{*})^2$ and is a price to pay for not assuming the layer-wise  positivity \citep{Lei2022}. If we further assume that all $\bm{P}_t^*$'s are non-negative definite, a slightly modified algorithm will exhibit consistency over the region complementary to the impossibility regime. Because we can replace the $\sum_{t=l-L+1}^{l}(\bm{P}_{t}^{*})^2$ with $\sum_{t=l-L+1}^{l}\bm{P}_{t}^{*}$ in \eqref{eq:def-proj} and \eqref{eq:def-eig}, which can be accurately estimated using $\sum_{t=l-L+1}^{l}\bm{A}_{t}$ \citep{Jing2020,Paul2020}. 
	\end{remark}
	
	\section{Numerical experiments}
	
	This section examines the numerical performance of the proposed method in various simulated examples and a real-life dynamic social media network dataset, and compare it against three existing methods, including a tensor subspace tracking approach \citep{Ozdemir2017}, a slightly modified spectral clustering method \citep{Cribben2017} and a minimum description length based method \citep{Cheung2020}. For simplicity, we denote our proposed method as SCP, its refined version as rSCP, and the three competing methods as TST, SCM, and MDL, respectively. The tuning parameters of TST, SCM and MDL are set as in $\widehat{\tau}_{k}<\tau_{k}^{*}$, \citet{Cribben2017} and \citet{Cheung2020}, whereas the tuning parameters of SCP  and rSCP are set following the theoretical results in Section 3. Particularly, we set $L=\lfloor T/20\rfloor$ and $b_{n,T}=\big(Ln\check{\rho}_{n,T}^{2}+L^{1/2}n\check{\rho}_{n,T}\sqrt{\max\{50,n\rho_{n,T}\}}\big)\log(n+T)/30$, where $\check{\rho}_{n,T}= \max \big\{\sum_{t=1}^{T}A_{ijt}/T,1\leq i <j \leq n\big\}$. 
	
	The numerical performance of each method is measured via its accuracy in estimating the number and locations of the true subspace change points. Specifically, for any change point set $\Gamma$ with cardinality $K$, its estimation accuracy is measured by $|K -K^{*}|$ and the Hausdorff distance. 
	
	\subsection{Simulation}
	\label{sec:simulation}
	
	In all simulated examples, a total of $T$ networks are generated from the dynamic stochastic block model with three change points, denoted as $\Gamma^{*} = \{sT,1/2T,3/4T\}$, where $s\in[0,1/4]$ controls the minimum distance $\Delta_{\min}^*$, and $\Delta_{\min}^*$ will be larger when $s$ increases. For each sub-interval, we generate $\bm{P}_{t}^{*} = \rho_{n,T}\bm{Z}^{(k)} \bm{B}_{t}(\bm{Z}^{(k)})^{\top}$ for $k=1,\ldots,4$, where $\bm{Z}^{(k)} \in \{0,1\}^{n\times R^k}$ represents the cluster membership matrix, $\bm{B}_{t} \in [0,1]^{R^k \times R^k}$ represents the connectivity matrix, and $\rho_{n,T}$ controls the sparsity of the network. To construct $\bm{Z}^{(1)}$, we randomly assign the $n$ nodes into three clusters, where each cluster consists of $0.3n$, $0.3n$, and $0.4n$ nodes, respectively. Next, we randomly choose $q$ percent of the $n$ nodes and reassign them to a different cluster to create $\bm{Z}^{(2)}$. To construct $\bm{Z}^{(3)}$, we remove the smallest cluster from $\bm{Z}^{(2)}$ and reassign its nodes into the remaining two clusters. Finally, we set $\bm{Z}^{(4)} = \bm{Z}^{(2)}$. For any $t \in [T]\setminus[\tau_{2}^{*},\tau_{3}^{*})$, we randomly choose $\bm{B}_t$ in $\{\bm{B}^1, \bm{B}^2\}$ with 
	\begin{align*}
		\bm{B}^{1} = \bm{W}_{1} \diag(1,0.5,0.5) \bm{W}_{1}^{T} \text{, and } \bm{B}^{2} =  \bm{W}_{2} \diag(1,0.5,-0.5) \bm{W}_{2}^{T},
	\end{align*}
	where 
	\begin{align*}
		\bm{W_1} =  \left[\begin{array}{ccc}
			3/4 & 1/4 & \sqrt{6}/4\\
			1/2& 3/4 & -\sqrt{6}/4  \\
			\sqrt{6}/4 & -\sqrt{6}/4 & -1/2 \\
		\end{array}\right]  \text{, and }
		\bm{W_2} =  \left[\begin{array}{ccc}
			1/2 & 1/2 & -\sqrt{2}/2\\
			1/2& 1/2 &\sqrt{2}/2  \\
			\sqrt{2}/2 & -\sqrt{2}/2 & 0 \\
		\end{array}\right]  .
	\end{align*}
	For $t \in [\tau_{2}^{*},\tau_{3}^{*})$, we randomly choose $\bm{B}_t$ from $(\bm{B}^1)_{[2],[2]}$ or $(\bm{B}^2)_{[2],[2]}$, which are the $2 \times 2$ upper left sub-matrices of $\bm{B}^1$ or $\bm{B}^2$. We consider three scenarios with various values of $\rho_{n,T}$, $q$ and $s$ to evaluate the performance of the proposed method in different aspects. 
	
	\textbf{Scenario I:} We fix $\rho_{n,T}=1$, $q = 0.5$, but vary the minimum distance $s$ in $\{1/10,1/15,1/20\}$.
	
	\textbf{Scenario II:} We fix $\rho_{n,T}=1$, $s = 1/4$, but vary the minimum change magnitude $q$ in $\{0.3,0.2,0.1\}$. 
	
	\textbf{Scenario III:} We fix $q=0.5$, $s = 1/4$, but vary network sparsity $\rho_{n,T}$ in $\{80/n,50/n,30/n\}$. 
	
	In each scenario, we present the averaged estimated change point numbers and the averaged Hausdorff distances obtained by different methods across 50 independent replications in Tables \ref{table:dist}-\ref{table:sparsity}. The symbol `/' is used to signify that a method fails to produce any results within 24 hours.
	
	\begin{table}[!htb]
		\caption{Averaged performance of different methods for estimating the number and locations of change points over 50 independent replications in Scenario I. The best performer is boldfaced.}
		\begin{subtable}[h]{1.0\textwidth}
			\caption{\footnotesize Number}
			\setlength{\tabcolsep}{12pt}
			\footnotesize
			\centering
			\begin{tabular}{c|c|c|c|c|c|c}
				\toprule[1pt]
				&$s$ & \multicolumn{1}{c|}{{SCP}} & \multicolumn{1}{c|}{{rSCP}} & \multicolumn{1}{c|}{{SCM}} & \multicolumn{1}{c|}{{TST}} & \multicolumn{1}{c}{{MDL}}  \\
				\midrule[1pt]
				\multirow{3}{*}{\makecell{$n=100$\\$T=200$}} & $1/10$ & $\bm{0.00(0.00)}$ &  $\bm{0.00(0.00)}$ &  $6.48(0.48)$ &  $0.98(0.12)$ & $0.80(0.12)$   \\
				& $1/15$ & $\bm{0.00(0.00)}$ &  $\bm{0.00(0.00)}$ &  $8.78(0.56)$ &  $1.16(0.12)$ & $0.52(0.08)$ \\
				& $1/20$ & $\bm{0.00(0.00)}$ &  $\bm{0.00(0.00)}$ &  $9.58(0.59)$ &  $1.20(0.15)$ & $0.84(0.10)$ \\
				\midrule[1pt]
				\multirow{3}{*}{\makecell{$n=200$\\$T=200$}} & $1/10$ & $\bm{0.00(0.00)}$ &  $\bm{0.00(0.00)}$ &  $5.98(0.50)$ &  $0.98(0.10)$ & $0.84(0.14)$   \\
				& $1/15$ & $\bm{0.00(0.00)}$ &  $\bm{0.00(0.00)}$ &  $8.70(0.54)$ &  $0.96(0.07)$ & $0.82(0.10)$ \\
				& $1/20$ & $\bm{0.00(0.00)}$ &  $\bm{0.00(0.00)}$ &  $9.80(0.57)$ &  $0.92(0.07)$ & $0.82(0.10)$ \\
				\midrule[1pt]
				\multirow{3}{*}{\makecell{$n=400$\\$T=300$}} & $1/10$ & $\bm{0.00(0.00)}$ &  $\bm{0.00(0.00)}$ &  / &  $0.92(0.08)$ & /   \\
				& $1/15$ & $\bm{0.00(0.00)}$ &  $\bm{0.00(0.00)}$ &  / &  $0.84(0.05)$ & / \\
				& $1/20$ & $\bm{0.00(0.00)}$ &  $\bm{0.00(0.00)}$ &  / &  $0.88(0.07)$ & / \\
				\bottomrule[1pt]                                                   
			\end{tabular}
		\end{subtable}
		
		\bigskip
		
		\begin{subtable}[h]{1.0\textwidth}
			\caption{\footnotesize Location}
			\setlength{\tabcolsep}{12pt}
			\footnotesize
			\centering
			\begin{tabular}{c|c|c|c|c|c|c}
				\toprule[1pt]
				&$s$ & \multicolumn{1}{c|}{{SCP}} & \multicolumn{1}{c|}{{rSCP}} & \multicolumn{1}{c|}{{SCM}} & \multicolumn{1}{c|}{{TST}} & \multicolumn{1}{c}{{MDL}}  \\
				\midrule[1pt]
				\multirow{3}{*}{\makecell{$n=100$\\$T=200$}} & $1/10$ & $2.56(0.08)$ &  $\bm{0.10(0.04)}$ &  $27.32(0.89)$ &  $28.36(1.73)$ & $15.82(1.62)$   \\
				& $1/15$ & $7.00(0.00)$ &  $\bm{0.38(0.14)}$ &  $28.94(1.00)$ &  $29.12(1.68)$ & $15.08(1.55)$ \\
				& $1/20$ & $10.00(0.00)$ &  $\bm{2.88(0.09)}$ &  $31.00(1.01)$ &  $29.54(1.70)$ & $15.36(1.44)$ \\
				\midrule[1pt]
				\multirow{3}{*}{\makecell{$n=200$\\$T=200$}} & $1/10$ & $1.74(0.06)$ &  $\bm{0.04(0.03)}$ &  $27.80(0.97)$ &  $32.52(1.66)$ & $21.82(1.94)$   \\
				& $1/15$ & $7.00(0.00)$ &  $\bm{0.08(0.04)}$ &  $30.60(0.93)$ &  $32.40(1.74)$ & $22.82(1.69)$ \\
				& $1/20$ & $10.00(0.00)$ &  $\bm{2.02(0.02)}$ &  $31.70(0.98)$ &  $34.00(1.62)$ & $21.40(2.32)$ \\
				\midrule[1pt]
				\multirow{3}{*}{\makecell{$n=400$\\$T=300$}} & $1/10$ & $1.28(0.06)$ &  $\bm{0.00(0.00)}$ &  / &  $54.44(3.02)$ & /   \\
				& $1/15$ & $10.00(0.00)$ &  $\bm{0.10(0.04)}$ &  / &  $55.88(2.82)$ & / \\
				& $1/20$ & $15.00(0.00)$ &  $\bm{2.00(0.00)}$ &  / &  $52.78(3.28)$ & / \\
				\bottomrule[1pt]                                                   
			\end{tabular}
			
		\end{subtable}
		\label{table:dist}
	\end{table}
	
	\begin{table}[!htb]
		\caption{Averaged performance of different methods for estimating the number and locations of change points over 50 independent replications in Scenario II. The best performer is boldfaced.}
		\begin{subtable}[h]{1.0\textwidth}
			\caption{\footnotesize Number}
			\setlength{\tabcolsep}{12pt}
			\footnotesize
			\centering
			\begin{tabular}{c|c|c|c|c|c|c}
				\toprule[1pt]
				&$q$ & \multicolumn{1}{c|}{{SCP}} & \multicolumn{1}{c|}{{rSCP}} & \multicolumn{1}{c|}{{SCM}} & \multicolumn{1}{c|}{{TST}} & \multicolumn{1}{c}{{MDL}}  \\
				\midrule[1pt]
				\multirow{3}{*}{\makecell{$n=100$\\$T=50$}} & $0.3$ & $\bm{0.00(0.00)}$ &  $\bm{0.00(0.00)}$ &  $2.48(0.11)$ &  $0.42(0.07)$ & $0.72(0.09)$   \\
				& $0.2$ & $\bm{0.00(0.00)}$ &  $\bm{0.00(0.00)}$ &  $2.50(0.09)$ &  $0.46(0.07)$ & $0.82(0.09)$ \\
				& $0.1$ & $\bm{0.02(0.02)}$ &  $\bm{0.02(0.02)}$ &  $2.78(0.07)$ &  $0.48(0.07)$ & $0.88(0.11)$ \\
				\midrule[1pt]
				\multirow{3}{*}{\makecell{$n=200$\\$T=100$}} & $0.3$ & $\bm{0.00(0.00)}$ &  $\bm{0.00(0.00)}$ &  $1.46(0.13)$ &  $0.86(0.05)$ & $0.72(0.08)$   \\
				& $0.2$ & $\bm{0.00(0.00)}$ &  $\bm{0.00(0.00)}$ &  $1.40(0.14)$ &  $0.64(0.07)$ & $0.82(0.13)$ \\
				& $0.1$ & $\bm{0.00(0.00)}$ &  $\bm{0.00(0.00)}$ &  $1.92(0.14)$ &  $0.84(0.05)$ & $0.52(0.09)$ \\
				\midrule[1pt]
				\multirow{3}{*}{\makecell{$n=100$\\$T=500$}} & $0.3$ & $\bm{0.00(0.00)}$ &  $\bm{0.00(0.00)}$ &  / &  $1.24(0.15)$ & /   \\
				& $0.2$ & $\bm{0.00(0.00)}$ &  $\bm{0.00(0.00)}$ &  / &  $1.26(0.17)$ & / \\
				& $0.1$ & $\bm{0.00(0.00)}$ &  $\bm{0.00(0.00)}$ &  / &  $1.26(0.14)$ & / \\
				\bottomrule[1pt]                                                   
			\end{tabular}
		\end{subtable}
		
		\bigskip
		
		\begin{subtable}[h]{1.0\textwidth}
			\caption{\footnotesize Location}
			\setlength{\tabcolsep}{12pt}
			\footnotesize
			\centering
			\begin{tabular}{c|c|c|c|c|c|c}
				\toprule[1pt]
				&$q$ & \multicolumn{1}{c|}{{SCP}} & \multicolumn{1}{c|}{{rSCP}} & \multicolumn{1}{c|}{{SCM}} & \multicolumn{1}{c|}{{TST}} & \multicolumn{1}{c}{{MDL}}  \\
				\midrule[1pt]
				\multirow{3}{*}{\makecell{$n=100$\\$T=50$}} & $0.3$ & $0.94(0.04)$ &  $\bm{0.16(0.05)}$ &  $20.44(0.88)$ &  $9.48(0.10)$ & $7.90(0.48)$   \\
				& $0.2$ & $1.04(0.04)$ &  $\bm{0.04(0.03)}$ &  $19.94(0.89)$ &  $9.54(0.07)$ & $7.86(0.53)$ \\
				& $0.1$ & $1.64(0.25)$ &  $\bm{0.46(0.32)}$ &  $22.88(0.65)$ &  $9.52(0.07)$ & $8.00(0.53)$ \\
				\midrule[1pt]
				\multirow{3}{*}{\makecell{$n=200$\\$T=100$}} & $0.3$ & $1.06(0.03)$ &  $\bm{0.02(0.02)}$ &  $25.30(1.36)$ &  $14.52(0.54)$ & $13.86(1.04)$   \\
				& $0.2$ & $1.18(0.05)$ &  $\bm{0.00(0.00)}$ &  $28.54(1.64)$ &  $13.62(0.57)$ & $13.12(0.95)$ \\
				& $0.1$ & $1.76(0.06)$ &  $\bm{0.00(0.00)}$ &  $32.82(1.90)$ &  $14.30(0.58)$ & $11.50(0.89)$ \\
				\midrule[1pt]
				\multirow{3}{*}{\makecell{$n=100$\\$T=500$}} & $0.3$ & $6.34(0.13)$ &  $\bm{0.02(0.02)}$ &  / &  $58.78(6.30)$ & /   \\
				& $0.2$ & $7.14(0.16)$ &  $\bm{0.02(0.02)}$ &  / &  $55.86(6.14)$ & / \\
				& $0.1$ & $9.02(0.18)$ &  $\bm{0.00(0.00)}$ &  / &  $54.76(5.92)$ & / \\
				\bottomrule[1pt]                                                   
			\end{tabular}
			
		\end{subtable}
		\label{table:jump}
	\end{table}
	
	\begin{table}[!htb]
		\caption{Averaged performance of different methods for estimating the number and locations of change points over 50 independent replications in Scenario III. The best performer is boldfaced.}
		\begin{subtable}[h]{1.0\textwidth}
			\caption{\footnotesize Number}
			\setlength{\tabcolsep}{12pt}
			\footnotesize
			\centering
			\begin{tabular}{c|c|c|c|c|c|c}
				\toprule[1pt]
				&$\rho_{n,T}$ & \multicolumn{1}{c|}{{SCP}} & \multicolumn{1}{c|}{{rSCP}} & \multicolumn{1}{c|}{{SCM}} & \multicolumn{1}{c|}{{TST}} & \multicolumn{1}{c}{{MDL}}  \\
				\midrule[1pt]
				\multirow{3}{*}{\makecell{$n=100$\\$T=150$}} & $80/n$ & $\bm{0.00(0.00)}$ &  $\bm{0.00(0.00)}$ &  $2.28(0.26)$ &  $1.08(0.13)$ & $1.36(0.17)$   \\
				& $50/n$ & $\bm{0.00(0.00)}$ &  $\bm{0.00(0.00)}$ &  $2.24(0.32)$ &  $2.40(0.17)$ & $1.06(0.15)$ \\
				& $30/n$ & $\bm{0.02(0.02)}$ &  $\bm{0.02(0.02)}$ &  $2.20(0.30)$ &  $4.24(0.15)$ & $1.28(0.14)$ \\
				\midrule[1pt]
				
				\multirow{3}{*}{\makecell{$n=100$\\$T=300$}} & $80/n$ & $\bm{0.00(0.00)}$ &  $\bm{0.00(0.00)}$ &  $3.96(0.41)$ &  $2.10(0.20)$ & $1.30(0.20)$   \\
				& $50/n$ & $\bm{0.00(0.00)}$ &  $\bm{0.00(0.00)}$ &  $4.18(0.42)$ &  $3.50(0.25)$ & $2.14(0.25)$ \\
				& $30/n$ & $\bm{0.00(0.00)}$ &  $\bm{0.00(0.00)}$ &  $3.82(0.42)$ &  $4.48(0.33)$ & $2.94(0.31)$ \\
				
				\midrule[1pt]
				\multirow{3}{*}{\makecell{$n=400$\\$T=400$}} & $80/n$ & $\bm{0.00(0.00)}$ &  $\bm{0.00(0.00)}$ &  / &  $1.54(0.18)$ & /   \\
				& $50/n$ & $\bm{0.00(0.00)}$ &  $\bm{0.00(0.00)}$ &  / &  $2.84(0.27)$ & / \\
				& $30/n$ & $\bm{0.00(0.00)}$ &  $\bm{0.00(0.00)}$ &  / &  $9.20(0.46)$ & / \\
				\bottomrule[1pt]                                                   
			\end{tabular}
		\end{subtable}
		
		\bigskip
		
		\begin{subtable}[!htb]{1.0\textwidth}
			\caption{\footnotesize Location}
			\setlength{\tabcolsep}{12pt}
			\footnotesize
			\centering
			\begin{tabular}{c|c|c|c|c|c|c}
				\toprule[1pt]
				&$\rho_{n,T}$ & \multicolumn{1}{c|}{{SCP}} & \multicolumn{1}{c|}{{rSCP}} & \multicolumn{1}{c|}{{SCM}} & \multicolumn{1}{c|}{{TST}} & \multicolumn{1}{c}{{MDL}}  \\
				\midrule[1pt]
				\multirow{3}{*}{\makecell{$n=100$\\$T=150$}} & $80/n$ & $2.36(0.07)$ &  $\bm{0.16(0.06)}$ &  $27.40(1.69)$ &  $24.20(1.32)$ & $15.88(1.20)$   \\
				& $50/n$ & $3.42(0.09)$ &  $\bm{0.40(0.09)}$ &  $30.04(1.33)$ &  $21.48(1.06)$ & $14.68(1.29)$ \\
				& $30/n$ & $5.54(0.41)$ &  $\bm{1.02(0.18)}$ &  $29.90(1.83)$ &  $23.66(0.82)$ & $15.20(1.14)$ \\
				\midrule[1pt]
				\multirow{3}{*}{\makecell{$n=100$\\$T=300$}} & $80/n$ & $4.08(0.07)$ &  $\bm{0.18(0.05)}$ &  $46.32(2.50)$ &  $27.50(1.69)$ & $18.64(1.68)$   \\
				& $50/n$ & $5.30(0.11)$ &  $\bm{0.26(0.07)}$ &  $45.08(2.33)$ &  $32.22(1.98)$ & $25.86(2.10)$ \\
				& $30/n$ & $7.40(0.14)$ &  $\bm{0.84(0.16)}$ &  $51.08(2.30)$ &  $36.58(2.22)$ & $29.34(2.54)$ \\
				
				\midrule[1pt]
				\multirow{3}{*}{\makecell{$n=400$\\$T=400$}} & $80/n$ & $6.42(0.14)$ &  $\bm{0.30(0.07)}$ &  / &  $24.78(2.22)$ & /   \\
				& $50/n$ & $9.26(0.15)$ &  $\bm{0.20(0.06)}$ &  / &  $35.16(2.71)$ & / \\
				& $30/n$ & $15.02(0.24)$ &  $\bm{0.48(0.09)}$ &  / &  $83.76(0.76)$ & / \\
				\bottomrule[1pt]                                                   
			\end{tabular}
			
		\end{subtable}
		\label{table:sparsity}
	\end{table}
	
	
	It is evident from Table \ref{table:dist}-\ref{table:sparsity} that SCP and rSCP perform significantly better than the other three methods in terms of both change point number and location recovery. More importantly, our refined estimate consistently outperforms the original one across all scenarios, exhibiting a significant improvement ranging from $72\%$ to $100\%$ in terms of location estimation errors. When $n$ and $T$ are held constant, we observe that the performance of SCP deteriorates as $s$, $q$, or $\rho_{n,T}$ decreases. This degradation can be attributed primarily to the increased difficulty of the problem.  Compared to  SCP, rSCP demonstrates greater resilience to variations in these three parameters in most scenarios, validating the robustness of our refined procedure. In Table \ref{table:dist}, we observe that rSCP clearly benefits from the increasing of node number $n$. In Table \ref{table:jump} and \ref{table:sparsity}, especially when $q$ or $\rho_{n,T}$ is extremely small, rSCP can locate the change point more accurately under larger $T$. This demonstrates the importance of effectively integrating the information across network layers and boosting the signal, which is largely overlooked by the other three comparison methods. Generally speaking, MDL performs slightly better than SCM and TST. However, its computational complexity is exceedingly high, resulting in computation times that are hundreds of times longer than our methods. 
	
	\subsection{UK politician social network}
	\label{sec:real_data}
	
	We now apply the proposed rSCP method to detect change points in the dynamic social networks of UK politicians \citep{Weaver2018}. It collects Twitter interactions among 648 UK politicians, including both Members of Parliament and British Members of the European Parliament, during December 12, 2014 to August 8, 2016. Each layer of network records whether the politicians interact with each other within a one-week interval, leading to a dynamic network with 85 layers.  Among the politicians, 516 of them were from the Conservative or Labour parties, who had connected with others at least once within the whole 85 weeks.
	
\begin{figure}[!htb]
		\flushleft
		\includegraphics[width=1\hsize]{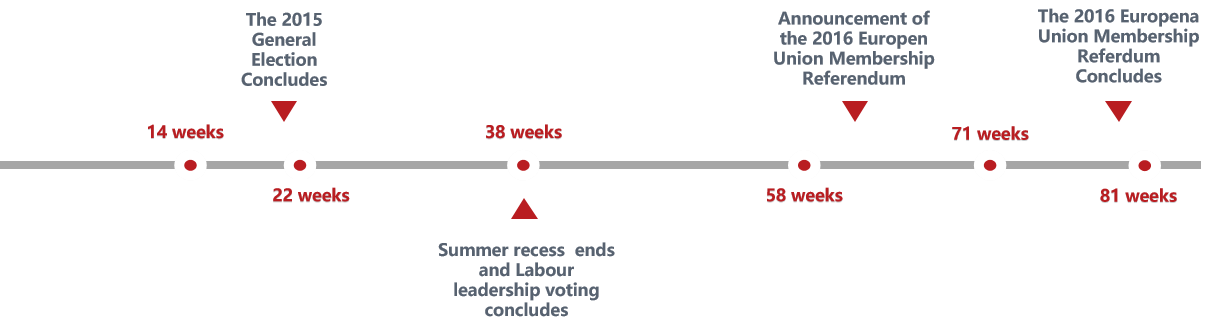}
		\caption{Six change points detected in the UK politician social networks}
		\label{fig:change_point}
	\end{figure}

	Our method detects six change points, summarized in Figure \ref{fig:change_point}. The second, third, fourth and sixth detected change points well align with the major political events: the 2015 General Election, the 2015 Labour Leadership Voting, and the announcement and conclusion of the 2016 European Union membership referendum (EU referendum). 
	
	It is also interesting to remark that the politicians do not interact much on Twitter before the first change point, and the rank of the estimated network subspace is just one. Soon after, as it is approaching the 2015 General Election day, the politicians leading left or right had started to interact within their communities. The rank of the estimated network subspace increases to two, and the estimated two communities agree with the political affiliations of politicians: $90\%$ of the members in the first community belong to the Conservative party, and the second community consists of Labour party members only. It is evident that the social separation between the Conservative and Labour party had become clearer after the first change point.
	
	The fifth change point marks the time that the 2016 European Union membership referendum had become a dominant topic among UK politicians. Before the fifth change point, the estimated two communities primarily reflects the conflicts between the two parties. More specifically, approximately $85\%$ of the members in the first community are affiliated with the Conservative party, while the second community consists of the Labour Party members only. Notice that only seven Labour Party members voted `Leave' in the EU referendum, and six of them are assigned to the first community, as the Conservative Party remains neutral during the EU referendum. As the referendum drawing became close, the three estimated communities are no longer associated with party affiliations, but significantly influenced by the politician's stance on the referendum. Within the first community, labeled as `Leave', $93.2\%$ of the politicians voted `Leave' and $5.1\%$ voted 'Remain'. Within the second community, labeled as `Remain', $9.2\%$ politicians voted `Leave' and $87.7\%$ voted `Remain'. The third community, labeled as `Wavering', interacts much more frequently with the other two communities rather than within itself. 
	
\begin{figure}[!htb]
		\centering
		\includegraphics[width = 0.5\hsize]{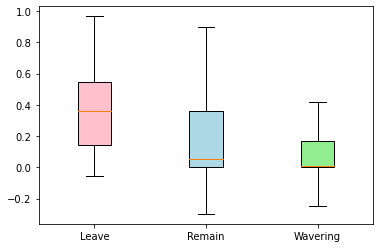}
		\caption{Boxplots of the Cohen's $\kappa$ indices between node pairs within each community.}
		\label{fig:cohen}
	\end{figure}

	
	As an interesting by-product, we observe that politicians within the `Leave' community exhibit more intense interactions among themselves compared with the other two communities. We calculate the average internal density \citep{Xu2022,Yang2012} for each community,
	\begin{align*}
		Acc_{k} = \sum_{\bm{c}_{i}=\bm{c}_{j}=k}\sum_{t=\widehat{\tau}_{5}}^{\widehat{\tau}_{6}-1}\frac{A_{t(i,j)}}{N_k(\widehat{\tau}_{6}-\widehat{\tau}_{5})},
	\end{align*}
	where $k \in [3]$, $\bm{c} \in [3]^{512}$ represents the cluster membership and $N_{k}$ denotes the total number of the $k$-th community. The average internal density of the `Leave' community is 0.6, which is notably higher than that of the other two communities, $0.27$ for `Remain' and $0.16$ for `Wavering'. Moreover, we also calculate Cohen's $\kappa$ index for each pair of the nodes under different snapshots, which is widely used to measure the similarity of node pairs \citep{Hoffman2015,Hoffman2018}. The indices are displayed in Figure \ref{fig:cohen}, revealing that politicians within the `Leave' community share more similar behaviors compared to the other two communities. Members of the `Leave' community display higher levels of engagement in providing reciprocal support on social media, which offers some interesting insights into their success in the referendum.
	
	\section{Conclusion}
	
	This article proposes a novel change point detection model for dynamic heterogeneous networks, which focuses on the alternation of the latent subspace but permits a continuous shift of the probability matrix. A sliding window algorithm with a refined procedure has been proposed to tackle this problem, where we define several novel statistics to quantify the subspace change. The effectiveness of our method is supported by various numerical experiments and asymptotic consistency in terms of change point detection. Particularly, the theoretical result can be established under nearly all parameters scaling for which this task is feasible. We believe this new change point detection model has broad application potential and can be extended to many more scenarios. One possible extension is to consider the node degree heterogeneity, which can continuously evolve over time.
	
	\section*{Appendix A: Technical proofs}
	
	\noindent \textbf{Proof of Lemma \ref{lemma:1}}.
	Note that
	\begin{align}
		\|\bm{UU}^{\top}-\bm{VV}^{\top}\|_{F}^2 & = \operatorname{tr}(\bm{UU}^{\top}+\bm{VV}^{\top}-\bm{UU}^{\top}\bm{VV}^{\top}-\bm{VV}^{\top}\bm{UU^{\top}}) \notag\\
		&= R_u+R_v-2\operatorname{tr}(\bm{V}^{\top}\bm{UU}^{\top}\bm{V}).
		\label{eq:subspace-distant-fnorm}
	\end{align}
	Then we have 
	\begin{align*}
		\|\bm{U}_{\perp}^{\top}\bm{V}\|_{F}^2  = \operatorname{tr}(\bm{V}^{\top}\bm{U}_{\perp}\bm{U}_{\perp}^{\top}\bm{V}) 
		& = \operatorname{tr}(\bm{V}^{\top}\bm{V})-\operatorname{tr}(\bm{V}^{\top}\bm{UU}^{\top}\bm{V}) \notag \\
		& = R_v-\big(R_u+R_v-	\|\bm{UU}^{\top}-\bm{VV}^{\top}\|_{F}^2\big)/2  \\
		& = \big(R_v-R_u+\|\bm{UU}^{\top}-\bm{VV}^{\top}\|_{F}^2\big)/2\notag,
	\end{align*}
	which directly leads to 
	\begin{align*}
		\|\bm{U}_{\perp}^{\top}\bm{V}\|_{2}^2 \geq \|\bm{U}_{\perp}^{\top}\bm{V}\|_{F}^2/R_{v} = \big(R_v-R_u+\|\bm{UU}^{\top}-\bm{VV}^{\top}\|_{F}^2\big)/(2R_{v}),
	\end{align*}
	where the first inequality follows from the assumption that $R_{v}\prec n-R_{u}$.  
	
	We then turn to bound $\|\bm{U_{\perp}}^{\top}\bm{VQV}^{\top}\|_{2}$. First, we have
	\begin{align}
		\|\bm{U_{\perp}}^{\top}\bm{VQV}^{\top}\|_{F}^2 &= \operatorname{tr}(\bm{U}_{\perp}^{\top}\bm{VQ}^2\bm{V}^{\top}\bm{U}_{\perp}) = \operatorname{tr}\big(\bm{O}\diag\big(\lambda_{1}\bm{(Q^2)},\cdots,\lambda_{R_v}(\bm{Q^2})\big) \bm{O}^{\top}\bm{V}^{\top}\bm{U}_{\perp}\bm{U}_{\perp}^{\top}\bm{V}\big) \notag \\
		& = \operatorname{tr}\big(\diag\big(\lambda_{1}\bm{(Q^2)},\cdots,\lambda_{R_v}(\bm{Q^2})\big) \bm{O}^{\top}\bm{V}^{\top}\bm{U}_{\perp}\bm{U}_{\perp}^{\top}\bm{VO}\big) \notag \\
		&\geq \lambda_{\min}(\bm{Q}^2)\operatorname{tr} \big(\bm{O}^{\top}\bm{V}^{\top}\bm{U}_{\perp}\bm{U}_{\perp}^{\top}\bm{VO}\big) =\lambda_{\min}(\bm{Q}^2)	\|\bm{U}_{\perp}^{\top}\bm{V}\|_{F}^2,
		\label{eq:lemma1-fnorm}
	\end{align}
	where $\bm{O}\diag\big(\lambda_{1}\bm{(Q^2)},\cdots,\lambda_{R_v}(\bm{Q^2})\big) \bm{O}^{\top}$ is the eigen-decomposition of $\bm{Q}^2$. It then follows that
	\begin{align*}
		\big\|{\bm U}_{\perp}^{\top} {\bm V} \bm{Q} \bm{V}^{\top}\big\|_{2} \geq R_{v}^{-1/2}\big\|{\bm U}_{\perp}^{\top} {\bm V} \bm{Q} \bm{V}^{\top}\big\|_{F}  \geq \sigma_{\min}\big(\bm{Q}\big)\sqrt{\big(R_v - R_u + \|\bm{U} \bm{U}^{\top}-\bm{V} \bm{V}^{\top}\|_{F}^2\big)/(2R_{v})}.
	\end{align*}
	This completes the proof of Lemma \ref{lemma:1}. \hfill $\blacksquare$
	
	\noindent\textbf{Proof of Proposition \ref{prop:noise-upper-bound}.}  First, it follows from the triangle inequality that
	\begin{align}
		&\ \Big\|\sum_{t=l-L+1}^{l}\big(\bm{A}_{t}^2-\bm{D}_{t}-(\bm{P}_{t}^{*})^2\big)\Big\|_2\notag \\
		= & \Big\|\sum_{t=l-L+1}^{l}\big((\bm{A}_{t}-\bm{P}_{t}^{*})^2+\bm{A}_{t}\bm{P}_t^{*}+\bm{P}_t^*\bm{A}_t-2(\bm{P}_{t}^{*})^2-\bm{D}_{t}\big)\Big\|_2\notag \\ 
		\leq &\  2\Big\|\sum_{t=l-L+1}^{l}(\bm{A}_{t}-\bm{P}_t^{*})\bm{P}_{t}^{*}\Big\|_2 +\Big\|\sum_{t=l-L+1}^{l}\diag\big((\bm{A}_{t}-\bm{P}_{t}^{*})^2\big)-\bm{D}_{t}\Big\|_2\notag \\
		&\hspace{11em}+\Big\|\sum_{t=l-L+1}^{l}(\bm{A}_{t}-\bm{P}_{t}^{*})^2-\diag\big((\bm{A}_{t}-\bm{P}_{t})^2\big)\Big\|_2.
		\label{eq:error-decompose}
	\end{align}
	Since $\alpha_{n}\preceq 1$, $\delta_{\min}^{*}\preceq 1$ and $ \Delta_{\min}^{*}\preceq L$, it then follows from Assumption \ref{assumption:signal-to-noise} that
	$n\rho_{n,T}\sqrt{L}\succeq n\rho_{n,T}\alpha_{n}^{2}\sqrt{\Delta_{\min}^{*}}\delta_{\min}^{*} \succ \log(n+T)$,
	when $\rho_{n,T}\preceq 1/n$. When $n\rho_{n,T}\succ 1$, we also have $n\rho_{n,T}\sqrt{L}\succ \max\{\sqrt{\Delta_{\min}^{*}},\big(\log(n+T)\big)^2/\sqrt{\Delta_{\min}^{*}}\}\succeq \log(n+T)$. Then, with probability at least $1-c_1(n+T)^{-2}$, it holds true that
	\begin{align*}
		\Big\|\sum_{t=l-L+1}^{l}(\bm{A}_{t}-\bm{P}_t^{*})\bm{P}_{t}^{*}\Big\|_2 & \leq \frac{c_2}{4}(n\rho_{n,T})^{3/2}\sqrt{L\log(n+T)}, \\
		\Big\|\sum_{t=l-L+1}^{l}\diag\big((\bm{A}_{t}-\bm{P}_{t})^2\big)-\bm{D}_{t}\Big\|_2 & \leq c_2Ln\rho_{n,T}^2, \ \mbox{and} \\
		\Big\|\sum_{t=l-L+1}^{l}(\bm{A}_{t}-\bm{P}_{t})^2-\diag\big((\bm{A}_{t}-\bm{P}_{t})^2\big)\Big\|_2 & \leq \frac{c_2}{4}n\rho_{n,T}(\sqrt{n\rho_{n,T}}+c_3)\sqrt{L\log(n+T)},
	\end{align*}
	where the first two inequalities directly follow from Appendix C of \citet{Lei2022}, and the last inequality follows from  (A.13)-(A.15) in Appendix B of \citet{Lei2022} with slight modification. The first desired inequality in Proposition \ref{prop:noise-upper-bound} then follows immediately.
	
	Next, we turn to derive the second inequality on $\big\|\sum_{t=l-L+1}^{l}(\bm{P}_{t}^{*})^2-\bar{\bm P}^{(l)}\big\|_{F}$, which can be upper bounded by $\sqrt{\rank\big(\sum_{t=l-L+1}^{l}(\bm{P}_{t}^{*})^2\big)+\rank\big(\bar{\bm P}^{(l)}\big)}\big\|\sum_{t=l-L+1}^{l}(\bm{P}_{t}^{*})^2-\bar{\bm P}^{(l)}\big\|_{2}$. Since $L<\Delta_{\min}^{*}$, there exists at most one change point in $[l-L+1,l]$ resulting in  $\rank\big(\sum_{t=l-L+1}^{l}(\bm{P}_{t}^{*})^2\big)\leq 2\max_{k \in [K^{*}+1]}R^{*(k)}$. Following from Weyl's inequality, with probability at least $1-c_1(n+T)^{-2}$, it holds true that
	\begin{align*}
		& \lambda_{2\max_{k \in [K^{*}+1]}R^{*(k)}+1}\Big(\sum_{t=l-L+1}^{l}\big(\bm{A}_{t}^2-\bm{D}_{t}\big)\Big)\\
		\leq\  & \lambda_{2\max_{k \in [K^{*}+1]}R^{*(k)}+1}\Big(\sum_{t=l-L+1}^{l}(\bm{P}_{t}^{*})^2\Big)+\Big\|\sum_{t=l-L+1}^{l}\big(\bm{A}_{t}^2-\bm{D}_{t}-(\bm{P}_{t}^{*})^2\big)\Big\|_2<b_{n,T},
	\end{align*}
	which further implies that the rank of $\bar{\bm P}^{(l)}$ is at most $2\max_{k \in [K^{*}+1]}R^{*(k)}$. Then, we have
	\begin{align*}
		&\Big\|\sum_{t=l-L+1}^{l}(\bm{P}_{t}^{*})^2-\bar{\bm P}^{(l)}\Big\|_{F} \leq 2\max_{k \in [K^{*}+1]}\sqrt{R^{*(k)}}	\Big\|\sum_{t=l-L+1}^{l}(\bm{P}_{t}^{*})^2-\bar{\bm P}^{(l)}\Big\|_{2} \\
		\leq \	& 2\max_{k \in [K^{*}+1]}\sqrt{R^{*(k)}}\bigg(\Big\|\sum_{t=l-L+1}^{l}\big(\bm{A}_{t}^2-\bm{D}_{t}\big)-\bar{\bm P}^{(l)}\Big\|_2+\Big\|\sum_{t=l-L}^{l}\big(\bm{A}_{t}^2-\bm{D}_{t}-(\bm{P}_{t}^{*})^2\big)\Big\|_2\bigg)\\
		\leq \	& 4\max_{k \in [K^{*}+1]}\sqrt{R^{*(k)}}b_{n,T}.
	\end{align*}
	This completes the proof of Proposition \ref{prop:noise-upper-bound}. \hfill $\blacksquare$
	
	\noindent\textbf{Proof of Theorem \ref{theorem:alg2}.} For $L \leq l \leq T-L+1 $, we consider the event 
	\begin{align*}
		\mathcal{B}_l = \bigg\{\Big\|\sum_{t=l-L+1}^{l}\big(\bm{A}_{t}^2-\bm{D}_{t}-(\bm{P}_{t}^{*})^2\big)&\Big\|_2 \leq b_{n,T}/\sqrt{\log(n+T)}, \\
		&\text{and } \Big\|\sum_{t=l-L+1}^{l}(\bm{P}_{t}^{*})^2-\bar{\bm P}^{(l)}\Big\|_{F}\leq  4\max_{k \in [K^{*}+1]}\sqrt{R^{*(k)}}b_{n,T}\bigg\}.
	\end{align*}
	By Proposition \ref{prop:noise-upper-bound}, we have $\mathbb{P}\big(\mathcal{B}_l\big) \geq 1-c_1(n+T)^{-2}$ for each $l$. Define the event $\mathcal{B}$ to be the intersection of $\{\mathcal{B}_l\}_{l=L}^{T-L+1}$, and then 
	\begin{align*}
		\mathbb{P}\big(\mathcal{B}\big)=1-\mathbb{P}\big(\mathcal{B}^c\big) = \mathbb{P}\big(\cup_{l\in [L,T-L+1]} \mathcal{B}_l^c\big) \geq 1-\sum_{l\in [L,T-L+1] } \mathbb{P}\big(\mathcal{B}_l^c\big) \geq 1-c_1(n+T)^{-1}.
	\end{align*}
	We then proceed to prove Theorem \ref{theorem:alg2} conditional on the event $\mathcal{B}$.
	
	For any $k \in [K^{*}]$, we first consider the scenario with $R^{*(k)}< R^{*(k+1)}$. Under event $\mathcal{B}$, Lemma 2 implies that $\widetilde \tau_{k} \geq \tau_{k}^{*}$ and $|\widetilde \tau_{k}-\tau_{k}^{*}|\prec L$. As a consequence, if $\widehat{\tau}_{k}\geq \tau_{k}^{*}$, we immediately have $\widehat{\tau}_{k}-\tau_{k}^{*}\leq \widetilde{\tau}_{k}-\tau_{k}^{*}\prec L$. If $\widehat{\tau}_{k}< \tau_{k}^{*}$, for any $l \in [\widetilde \tau_{k}-L+1,\tau_{k}^{*}]$, it can be shown conditional on $\mathcal{B}$ that 
	\begin{align*}
		\lambda_{R^{*(k)}}\bigg(\sum_{t=l-L}^{l-1}\bm{A}_{t}^2-\bm{D}_{t}\bigg) & \geq \lambda_{R^{*(k)}}\bigg(\sum_{t=l-L}^{l-1}(\bm{P}_{t}^{*})^2\bigg)-\bigg\|\sum_{t=l-L}^{l-1}\big(\bm{A}_{t}^2-\bm{D}_{t}-(\bm{P}_{t}^{*})^2\big)\bigg\|_2  \\
		&\geq Ln^2\alpha_{n}^{2}\rho_{n,T}^2-b_{n,T}/\sqrt{\log(n+T)}, 
	\end{align*}
	where the second inequality follows from Assumption \ref{assumption:balanced-singularvalue} and Proposition \ref{prop:noise-upper-bound}. Further, 
	\begin{align}
		\frac{b_{n,T}}{Ln^2\alpha_{n}^2\rho_{n,T}^{2}}&=\frac{c_2\left(L n \rho_{n, T}^2+\sqrt{L} n \rho_{n, T} \max \left\{\sqrt{n \rho_{n, T}}, c_3\right\}\right) \log (n+T)}{Ln^2\alpha_{n}^2\rho_{n,T}^2}\notag\\
		& \preceq \frac{\sqrt{L}n\rho_{n,T}\log(n+T)}{Ln^2\alpha_{n}^2\rho_{n,T}^{2}}+\frac{\sqrt{L}n^{3/2}\rho_{n,T}^{3/2}\log(n+T)}{Ln^2\alpha_{n}^2\rho_{n,T}^{2}}\notag \\
		&\preceq \frac{\log(n+T)}{\sqrt{\Delta_{\min}^{*}}n\rho_{n,T}\alpha_n^2}+\frac{\log(n+T)}{\sqrt{\Delta_{\min}^{*}n\rho_{n,T}}\alpha_n^2} \prec 1,
		\label{eq:noise-smaller-signal}
	\end{align}
	where the first inequality follows from $Ln\rho_{n,T}^2\preceq \sqrt{L}n\rho_{n,T}\max\{\sqrt{n\rho_{n,T}},c_3\}$ when $n\succeq T$, and the second inequality is due to the fact that $\Delta_{\min}^{*}\preceq L \le \Delta_{\min}^{*}/3$, and the last inequality follows from Assumption \ref{assumption:signal-to-noise} and the fact that $\delta_{\min}^{*}\preceq 1$. Therefore, we have $\lambda_{R^{*(k)}}\Big(\sum_{t=l-L}^{l-1}\bm{A}_{t}^2-\bm{D}_{t}\Big) \succ b_{n,T}$. However, conditional on event $\mathcal{B}$, we also have 
	\begin{align*}
		\lambda_{R^{*(k)}+1}\bigg(\sum_{t=l-L}^{l-1}\bm{A}_{t}^2-\bm{D}_{t}\bigg) \leq \lambda_{R^{*(k)}+1}\bigg(\sum_{t=l-L}^{l-1}(\bm{P}_{t}^{*})^2\bigg)+\bigg\|\sum_{t=l-L}^{l-1}\big(\bm{A}_{t}^2-\bm{D}_{t}-(\bm{P}_{t}^{*})^2\big)\bigg\|_2 \prec b_{n,T}. 
	\end{align*}
	It follows immediately that $\rank \big ( \widehat{\bm{U}}^{(l-1)} \big ) = R^{*(k)}$. By the variant of Davis-Khan Theorem \citep{Yu2015} , we further have
	\begin{align*}
		\big\|\widehat{\bm{U}}_{\perp}^{(l-1)}(\widehat{\bm{U}}_{\perp}^{(l-1)})^{\top} & -\bm{V}_{\perp}^{*(k)}(\bm{V}_{\perp}^{*(k)})^{\top}\big\|_F =\sqrt{2}\big\|(\widehat{\bm U}_{\perp}^{(l-1)})^{\top}\bm{V}^{*(k)}\big\|_F  \\
		&\leq  \frac{2\sqrt{2R^{*(k)}}\Big\|\sum_{t=l-L}^{l-1}\big(\bm{A}_{t}^2-\bm{D}_{t}-(\bm{P}_{t}^{*})^2\big)\Big\|_2}{\lambda_{R^{*(k)}}\big(\sum_{t=l-L}^{l-1}(\bm{P}_{t}^{*})^2\big)} \prec \frac{b_{n,T}}{Ln^2\alpha_{n}^2\rho_{n,T}^2}.
	\end{align*}
	
	Let $\Pi_{ref1}(l) = \tr\Big(\bm{U}_{\perp}^{*(l-1)}(\bm{U}_{\perp}^{*(l-1)})^{\top}\sum_{t=l}^{l+L-1}(\bm{P}_{t}^{*})^2\Big)$, where  $\bm{U}^{*(l-1)}$ and $\bm{U}^{*(l+L-1)}$ denotes the eigenvectors corresponding to the nonzero eigenvalues of $\sum_{t=l-L}^{l-1}(\bm{P}_{t}^{*})^2$.
	Then, we can establish an upper bound on the difference between $\widehat \Pi_{ref1}(l)$ and $\Pi_{ref1}(l)$ by
	\begin{align*}
		&\big|\widehat \Pi_{ref1}(l)-\Pi_{ref1}(l)\big| \notag \\
		= \ &\Big| \tr\big((\widehat{\bm U}_{\perp}^{(l-1)})^{\top}\bar{\bm P}^{(l+L-1)} \widehat{\bm U}_{\perp}^{(l-1)}\big)-\tr\big((\bm V_{\perp}^{*(k)})^{\top}\sum_{t=l}^{l+L-1}(\bm P_{t}^{*})^{2}\bm V_{\perp}^{*(k)}\big)\Big| \notag\\
		=\ & \Big| \tr\big( (\widehat{\bm U}_{\perp}^{(l-1)})^{\top}\big(\bar{\bm P}^{(l+L-1)}-\sum_{t=l}^{l+L-1}(\bm P_{t}^{*})^{2}\big)\widehat{\bm U}_{\perp}^{(l-1)}\big)+\tr\big((\widehat{\bm{U}}_{\perp}^{(l-1)}(\widehat{\bm{U}}_{\perp}^{(l-1)})^{\top}-\bm{V}_{\perp}^{*(k)}(\bm{V}_{\perp}^{*(k)})^{\top})\sum_{t=l}^{l+L-1}(\bm P_{t}^{*})^2\big)\Big|\notag\\
		\leq \ &  2\max_{k\in K^{*}}R^{*(k)}\Big(\big\|\bar{\bm P}^{(l+L-1)}-\sum_{t=l}^{l+L-1}(\bm P_{t}^{*})^{2}\big\|_2+\big\|\widehat{\bm{U}}_{\perp}^{(l-1)}(\widehat{\bm{U}}_{\perp}^{(l-1)})^{\top}-\bm{V}_{\perp}^{*(k)}(\bm{V}_{\perp}^{*(k)})^{\top}\big\|_2\big\|\sum_{t=l}^{l+L-1}(\bm P_{t}^{*})^2\big\|_2\Big)\notag \\
		\preceq\ & b_{n,T},
	\end{align*}
	where the first inequality follows from the fact that $\max\big\{\rank\big(\bar{\bm P}^{(l+L-1)}\big),\rank\big(\sum_{t=l}^{l+L-1}(\bm P_{t}^{*})^2\big)\big\}\leq 2\max_{k\in K^{*}}R^{*(k)}$, and the last inequality is a direct result from $\big\|\bar{\bm P}^{(l+L-1)}-\sum_{t=l}^{l+L-1}(\bm P_{t}^{*})^{2}\big\|_2\preceq b_{n,T}$ given the event $\mathcal{B}$ and the fact that
	\begin{align*}
		\big\|\widehat{\bm{U}}_{\perp}^{(l-1)}(\widehat{\bm{U}}_{\perp}^{(l-1)})^{\top} & -\bm{V}_{\perp}^{*(k)}(\bm{V}_{\perp}^{*(k)})^{\top}\big\|_2\big\|\sum_{t=l}^{l+L-1}(\bm P_{t}^{*})^2\big\|_2 \\
		\preceq & Ln^2\alpha_{n}^2\rho_{n,T}^2 \big\|\widehat{\bm{U}}_{\perp}^{(l-1)}(\widehat{\bm{U}}_{\perp}^{(l-1)})^{\top}-\bm{V}_{\perp}^{*(k)}(\bm{V}_{\perp}^{*(k)})^{\top}\big\|_F \prec b_{n,T}.
	\end{align*}
	Moreover, we have
	\begin{align}
		\Pi_{ref1}(\tau_{k}^{*})-\Pi_{ref1}(\widehat{\tau}_{k}) \leq \Pi_{ref1}(\tau_{k}^{*})-\widehat{\Pi}_{ref1}({\tau}_{k}^{*})+\widehat{\Pi}_{ref1}(\widehat{\tau}_{k})-\Pi_{ref1}(\widehat{\tau}_{k}) \preceq b_{n,T}.
		\label{eq:refined-err-upperbound}
	\end{align}
	Note that $R^{*(k+1)}-R^{*(k)}+\delta^{*(k+1)}\geq 1$ when $R^{*(k+1)}>R^{*(k)}$. For any $l \in [\widetilde{\tau}_{k}-L+1,\tau_{k}^{*})$, the difference between $\Pi_{ref1}(\tau_{k}^{*})$ and $\Pi_{ref1}(l)$ can be lower bounded by
	\begin{align}
		&  \Pi_{ref1}(\tau_{k}^{*})-\Pi_{ref1}(l) = \tr\Big(\bm{V}_{\perp}^{*(k)}(\bm{V}_{\perp}^{*(k)})^{\top}\sum_{t=\tau_{k}^{*}}^{\tau_{k}^{*}+L-1}(\bm{P}_{t}^{*})^2\Big)-\tr\Big(\bm{V}_{\perp}^{*(k)}(\bm{V}_{\perp}^{*(k)})^{\top}\sum_{t=\tau_{k}}^{l+L-1}(\bm{P}_{t}^{*})^2\Big) \notag\\
		= & \tr\Big((\bm{V}_{\perp}^{*(k)})^{\top}\sum_{t=l+L}^{\tau_{k}^{*}+L-1}(\bm{P}_{t}^{*})^2 \bm{V}_{\perp}^{*(k)}\Big) = \rho_{n,T}^2\Big\|(\bm{V}_{\perp}^{*(k)})^{\top}\bm{V}^{*(k+1)}\Big ( \sum_{t=l+L}^{\tau_{k}^{*}+L-1}\bm{M}_{t}^2 \Big )^{1/2}(\bm{V}^{*(k+1)})^{\top}\Big\|_{F}^2 \notag\\
		\geq & \frac{\rho_{n,T}^2\lambda_{\min}\Big(\sum_{t=l+L}^{\tau_{k}^{*}+L-1}(\bm{M}_{t}^{*})^2\Big)(R^{*(k+1)}-R^{*(k)}+\delta^{*(k)})}{2}  \succeq (\tau_{k}^{*}-l)n^2\alpha_{n}^2\rho_{n,T}^2,
		\label{eq:refined-err-lowerbound}
	\end{align}
	where the first inequality follows from Lemma \ref{lemma:1}. When $\widehat{\tau}_{k}<\tau_{k}^{*}$, combining \eqref{eq:refined-err-upperbound} and \eqref{eq:refined-err-lowerbound} yields that
	\begin{align*}
		\tau_{k}^{*}-\widehat{\tau}_{k} \preceq \frac{b_{n,T}}{n^2\alpha_{n}^2\rho_{n,T}^2} \prec \frac{Ln^2\alpha_{n}^2\rho_{n,T}^2\sqrt{\delta_{\min}^{*}}}{n^2\alpha_{n}^2\rho_{n,T}^2} \preceq L,
	\end{align*}
	where the second inequality follows from a similar treatment in deriving \eqref{eq:noise-smaller-signal}. 
	
	When $R^{*(k)}> R^{*(k+1)}$, we have $\widetilde{\tau}_{k}\leq \tau_{k}^{*}$ and $|\widetilde{\tau}_{k}-\tau_{k}^{*}|\prec L$. Similar to the above analysis, we can prove that $|\widehat{\tau}_{k}-\tau_{k}^{*}|\prec L$. When $R^{*(k)} = R^{*(k+1)}$, following the proof of Lemma 2, $\check \tau_{k}$ is a consistent estimate of $\tau_{k}^{*}$ and locates on the left side of $\tau_{k}^{*}$ with high probability. It is clear that the estimation error of $\widehat{\tau}_{k}$ is controlled by $\max\big\{|\check{\tau}_{k}-\tau_{k}^{*}|,|\widetilde \tau_{k}-\tau_{k}^{*}|\big\}$. This completes the proof of Theorem \ref{theorem:alg2}. \hfill $\blacksquare$

	\noindent\textbf{Proof of Theorem \ref{theorem:mini-max}.} We prove Theorem \ref{theorem:mini-max} via Le Cam's method. First, we construct two sequences of probability matrices $\{\bm{P}_{t}^{(1)}\}_{t=1}^{T}$ and $\{\bm{P}_{t}^{(2)}\}_{t=1}^{T}$, which share the same network sparsity $\bar{\rho}_{n,T}$ and minimum distance $\bar{\Delta}_{\min}^{*}$ between adjacent change points, under the condition that $n\bar{\rho}_{n,T}\bar{\Delta}_{\min}^{*}\simeq \big(\log(n+T)\big)^{-1}$. Particularly, for any $t \in [T]\setminus[\bar{\Delta}_{\min}^{*}]$, we let $\bm{P}_{t}^{(1)}=\widetilde{\bm P}=\bar{\rho}_{n,T}\widetilde{\bm Z}\widetilde{\bm B}\widetilde{\bm Z}^{\top}$, where $\widetilde{\bm Z}\in \{0,1\}^{n\times m^{(1)}}$ with $m^{(1)}>1$ denotes the community membership matrix, $\mathcal{N}_m = \{i: \widetilde {\bm Z}_{im} = 1\}$ denotes the $m$-th community with $\min_{m\in[m^{(1)}]} |\mathcal{N}_m | \simeq n$, and $\widetilde{\bm B} \in [0,1]^{m^{(1)} \times m^{(1)}}$ denotes the connecting probability matrix, with $\widetilde{\bm B}_{-m^{(1)},-m^{(1)}} = \frac{m^{(1)}-1}{m^{(1)}}\diag(\bm{1}_{m^{(1)}-1})+\frac{1}{m^{(1)}}\bm{1}_{m^{(1)}-1}\bm{1}_{m^{(1)}-1}^{\top}$ and $\widetilde{\bm B}_{m^{(1)}\cdot}=e_{m^{(1)}}$. For any $t\in [\bar{\Delta}_{\min}^{*}]$, we let $\bm{P}_{t}^{(1)}= \bm{P}^{(1)} = \bar{\rho}_{n,T}\bm{Z}^{(1)}\bm{B}(\bm{Z}^{(1)})^{\top}$, where $\bm{B}= \bigg(\begin{array}{cc}
		\widetilde{\bm B}_{-m^{(1)},-m^{(1)}} & 0 \\
		0 &  \frac{4}{3}\bm{M} 
	\end{array}\bigg)$ with $\bm{M} = (\bm{I}_{2}+\bm{1}_{2}\bm{1}_{2}^{\top})/2$, and $\bm{Z}^{(1)} \in \{0,1\}^{n\times (m^{(1)}+1)}$ is defined as
	\begin{align*}
		(\bm{Z}_{i \cdot}^{(1)})^{\top} = \left\{\begin{array}{ll}
			b_{i}^{(1)}(\widetilde{\bm Z}_{i\cdot},0)^{\top}+(1-b_{i}^{(1)})\bm{e}_{m^{(1)}+1}(m^{(1)}+1),   & \text{if } i\in \mathcal{N}_{m^{(1)}};   \\
			(\widetilde{\bm Z}_{i\cdot},0)^{\top},
			&  \text{Otherwise},
		\end{array}\right.
	\end{align*}
	with $b_{i}^{(1)}\sim\text{Bern}(0.5)$. On the other hand, we let $\bm{P}_{t}^{(2)} = \widetilde{\bm P}$ for $t \in [T-\bar{\Delta}_{\min}^{*}]$, and set $\bm{P}_{t}^{(2)}=\bm{P}^{(2)}$ for $t\in[T]\setminus [T-\bar{\Delta}_{\min}^{*}]$, with $\bm{P}^{(2)}$ generated similarly as $\bm{P}^{(1)}$. Further, let $f_1$ and $f_2$ denote the multivariate Bernoulli distributions, with parameters $\{\bm{P}_{t}^{(1)}\}_{t=1}^{T}$ and $\{\bm{P}_{t}^{(2)}\}_{t=1}^{T}$, respectively. 
	
	We then show that both $f_1$ and $f_2$ are contained in $\mathcal{Q}$ with high probability. Denote $\widetilde{\bm V} = \widetilde{\bm Z}\diag(\widetilde{\bm Z}^{\top}\bm{1}_{n})^{-1/2}$, $\widetilde{\bm{M}}=\diag(\widetilde{\bm Z}^{\top}\bm{1}_{n})^{1/2}\widetilde{\bm B}\diag(\widetilde{\bm Z}^{\top}\bm{1}_{n})^{1/2}$, $\bm V^{(1)} = {\bm Z}^{(1)}\diag\big(({\bm Z}^{(1)})^{\top}\bm{1}_{n}\big)^{-1/2}$, and $\bm{M}^{(1)}=\diag\big(({\bm Z}^{(1)})^{\top}\bm{1}_{n}\big)^{1/2}{\bm B}\diag\big(({\bm Z}^{(1)})^{\top}\bm{1}_{n}\big)^{-1/2}$, then we have $\widetilde{\bm P} = \bar{\rho}_{n,T}\widetilde{\bm V}\widetilde{\bm M}\widetilde{\bm V}^{\top}$ and $\bm{P}^{(1)} = \bar{\rho}_{n,T}\bm{V}^{(1)}\bm{M}^{(1)}(\bm{V}^{(1)})^{\top}$. The singular values of $\widetilde{\bm M}$ can be bounded by
	\begin{align*}
		\sigma_{\max}(\widetilde{\bm M}) &= \big\|\diag(\widetilde{\bm Z}^{\top}\bm{1}_{n})^{1/2}\widetilde{\bm B}\diag(\widetilde{\bm Z}^{\top}\bm{1}_{n})^{1/2}\big\|_{2} \leq \max_{m \in [m^{(1)}]}|\mathcal{N}_{m}|\sigma_{\max}(\widetilde{\bm B}) \simeq n, \\
		\sigma_{\min}(\widetilde{\bm M}) &= \big\|\diag(\widetilde{\bm Z}^{\top}\bm{1}_{n})^{-1/2}\widetilde{\bm B}^{-1}\diag(\widetilde{\bm Z}^{\top}\bm{1}_{n})^{-1/2}\big\|_{2}^{-1}
		\geq \big(\max_{m \in [m^{(1)}]}\title{n}_{m}^{-1}\|\widetilde{\bm B}^{-1}\|_{2}\big)^{-1}\\
		&=\min_{m \in [m^{(1)}]}|\mathcal{N}_{m}|\sigma_{\min}(\widetilde{\bm B})\simeq n,
	\end{align*}
	where $\sigma_{\min}(\widetilde{\bm B})\geq (m^{(1)})^{-1}$ follows from the Gershgorin Cicle Theorem. Further, as $\bm V^{(1)}$ is an orthonormal matrix, we have $\sigma_{\max}(\bm{M}^{(1)})=\sigma_{1}(\bm{P}^{(1)})/\bar{\rho}_{n,T}$ and $\sigma_{\min}(\bm{M}^{(1)})=\sigma_{m^{(1)}+1}(\bm{P}^{(1)})/\bar{\rho}_{n,T}$. Therefore, with probability at least $1-n^{-1}$, it follows from Lemma 3 that
	\begin{align*}
		\sigma_{\max}(\bm{M}^{(1)})\simeq \sigma_{\min}(\bm{M}^{(1)})\simeq \sigma_{\max}(\widetilde{\bm M})\simeq \sigma_{\min}(\widetilde{\bm M}) \simeq n, 
	\end{align*}
	implying that $f_{1}$ satisfies Assumption \ref{assumption:balanced-singularvalue} with $\alpha_{n}^{(1)}\simeq 1$. Additionally,  the magnitude of the minimum subspace change $\delta_{\min}^{*(1)}$ does not vanish with $n$ since 
		\begin{align*}
			\delta_{\min}^{*(1)} &= \|\widetilde{\bm V}\widetilde{\bm V}^{\top}-\bm{V}^{(1)}(\bm V^{(1)})^{\top}\|_{F}^2 = \rank(\widetilde{\bm V})+\rank(\bm{V}^{(1)})-\tr(\widetilde{\bm V}^{\top}\bm V^{(1)}(\bm V^{(1)})^{\top}\widetilde{\bm V})\\
			&\geq 2m^{(1)}+1-2m^{(1)} =1,
		\end{align*}
		where the second equality follows from \eqref{eq:subspace-distant-fnorm}. Therefore, $n\bar{\rho}_{n,T}^{(1)}(\alpha_{n}^{(1)})^{4}\bar{\Delta}_{\min}^{*}\delta_{\min}^{*(1)}\simeq \big(\log(n+T)\big)^{-1}$, and thus $f_{1}$ belongs to $\mathcal{Q}$ with probability at least $1-n^{-1}$. A similar treatment also yields that $f_{2}$ belongs to $\mathcal{Q}$ with probability at least $1-n^{-1}$. 
	
	Next, let $\phi$ be a map from the observed adjacency networks to $\{1,2\}$, then there exists a constant $c_5$ such that 
	\begin{align}
		& \inf_{\check{\Gamma}}\sup_{Q \in \mathcal{Q}} \mathbb{E}_{Q}\Big(H\big(\check{\Gamma},\Gamma(Q)\big)\Big) \notag \\
		\geq \ &\sup_{\{Q^{1},Q^{2}\}\subset \mathcal{Q}} \frac{\Big(H\big(\Gamma(Q^{1}),\Gamma(Q^{2})\big)\Big)}{4}\inf_{\phi}\Big(Q^{1}(\phi\big(\{\bm A_{t}\}_{t=1}^{T}\big) \neq 1)+ Q^{2}(\phi\big(\{\bm A_{t}\}_{t=1}^{T}\big) \neq 2)\Big)\notag \\
		\geq \ &  \frac{\Big(H\big(\Gamma(f_{1}),\Gamma(f_{2})\big)\Big)}{4}\inf_{\phi}\Big(f_{1}(\phi\big(\{\bm A_{t}\}_{t=1}^{T}\big) \neq 1)+ f_{2}(\phi\big(\{\bm A_{t}\}_{t=1}^{T}\big) \neq 2)\Big)-c_5n^{-1}  \notag \\
		= \ &  \frac{\Big(H\big(\Gamma(f_{1}),\Gamma(f_{2})\big)\Big)}{4}\big(1-\|f_{1}-f_{2}\|_{TV}\big)-c_5n^{-1},
		\label{eq:minimax-transform}
	\end{align}
	where $\|\cdot\|_{TV}$ stands for the total variation distance, the first inequality follows from Proposition 15.1 in \citet{Wainwright2019}, the second inequality follows from C.23 in \citet{Jin2022}, and the last equality follows from equation (15.13) in \citet{Wainwright2019}.  Let $f_{0}$ be a multivariate Bernoulli distribution with parameter $\widetilde{\bm P}$ across all $T$ snapshots. As $f_{0}$ and $f_{1}$ only differ at $[\bar{\Delta}_{\min}^{*}]$ snapshots and $n\bar{\rho}_{n,T}\bar{\Delta}_{\min}^{*}\prec 1$, it follows from Lemma 4 that $\chi^{2}(f_{0},f_{1})\prec 1$. Similarly, we also have $\chi^{2}(f_{0},f_{2})\prec 1$. Therefore, for any $0<\bar{\Delta}_{\min}^{*}<T/4$, it holds true that
	\begin{align*}
		\frac{H\big(\Gamma(f_{1}),\Gamma(f_{2})\big)}{4}\big(1-\|f_{1}-f_{2}\|_{TV}\big) & \geq \frac{T-2\bar{\Delta}_{\min}^{*}}{2}\big(1-\|f_{1}-f_{0}\|_{TV}-\|f_{1}-f_{0}\|_{TV}\big)\\
		&\geq \frac{\bar{\Delta}_{\min}^{*}}{2}\big(1-\sqrt{\chi^{2}(f_{0},f_{1})}-\sqrt{\chi^{2}(f_{0},f_{2})}\big)\geq \frac{\bar{\Delta}_{\min}^{*}}{4}, 
	\end{align*}
	when $n$ is sufficiently large. The desired lower bound then follows immediately with $c_4 < 1/4$. \hfill $\blacksquare$

		\newpage
	\bibliographystyle{apa}
	\bibliography{DNSCPD}
	
\end{document}